\documentclass[aps,prl,floats,superscriptaddress,showpacs,twocolumn]{revtex4-1}

\usepackage{amsmath, amsthm, amssymb}
\usepackage{enumitem}
\usepackage{graphics,graphicx}
\usepackage{epsfig}
\usepackage{times}
\usepackage{dcolumn}
\usepackage{bm}

\usepackage{ulem, color}

\begin{document}

\title{Synchronization of Intermittent Behavior in Ensembles of Multistable Dynamical Systems}

\author{R. Sevilla-Escoboza}
\affiliation{Centro Universitario de los Lagos, Universidad de Guadalajara, Enrique D\'{i}az de Leon, Paseos de la Monta\~na, Lagos de Moreno, Jalisco 47460, Mexico}
\author{J. M. Buld\'u}
\affiliation{Laboratory of Biological Networks, Center for Biomedical Technology, Technical University of Madrid, Pozuelo de Alarc\'{o}n, 28223 Madrid, Spain}
\affiliation{Complex Systems Group, Universidad Rey Juan Carlos, 28933 M\'ostoles, Madrid, Spain}
\author{A. N. Pisarchik}
\affiliation{Computational Systems Biology Group, Center for Biomedical Technology, Technical University of Madrid, Pozuelo de Alarc\'{o}n, 28223 Madrid, Spain}
\affiliation{Centro de Investigaciones en Optica, Loma del
  Bosque 115, 37150 Leon, Guanajuato, Mexico}
\author{S. Boccaletti}
\affiliation{CNR-Istituto dei Sistemi Complessi, Via Madonna del Piano, 10, 50019 Sesto Fiorentino, Italy}
\affiliation{The Italian Embassy in Israel, 25 Hamered Street, 68125 Tel Aviv, Israel}
\author{R. Guti\'errez}
\affiliation{Department of Chemical Physics, The Weizmann Institute of Science, Rehovot 76100, Israel}

\date{\today}

\begin{abstract}
We propose a methodology to analyze synchronization in an ensemble of diffusively coupled multistable systems. First, we study how two bidirectionally coupled multistable oscillators synchronize and demonstrate  the high complexity of the basins of attraction of coexisting synchronous states. Then, we propose the use of the Master Stability Function (MSF) for multistable systems to describe synchronizability, even during intermittent behaviour, of a network of multistable oscillators, regardless of both the number of coupled oscillators and the interaction structure. In particular, we show that a network of multistable elements is synchronizable for a given range of topology spectra and coupling strengths, irrespective of specific attractor dynamics to which different oscillators are locked, and even in the presence of intermittency. Finally, we experimentally demonstrate the feasibility and robustness of the MSF approach with a network of multistable electronic circuits. 
\end{abstract} 

\pacs{89.75.Hc,89.75.Fb}
\maketitle

\section{Introduction}
During the last two decades great progress has been made in understanding synchronization of chaotic systems \cite{pikovsky2001,boccaletti2002}. Nevertheless, synchrony in dissipative dynamical systems with coexisting attractors remains relatively unexplored and poorly understood to this very day. This relative lack of activity is hard to reconcile with the fact that multistability has been  observed in numerous nonlinear systems in many fields of science, such as laser physics \cite{arecchi1982}, neuroscience \cite{park1999,schwartz2012}, cardiac dynamics \cite{Comtois05}, genetics \cite{Ullner08, Vetsigiana09}, cell signaling \cite{Angeli04}, and ecology \cite{Horan11} amongst others; moreover, in situations where synchronization is actually a collective behavior known to play a primary role. Even forms of extreme multistability, i.e. the coexistence of infinitely many attractors in phase space, have been recently observed in experiments \cite{Patel14}. Many of these results as well as some known coupling mechanisms and dynamical phenomena that seem to be correlated to the emergence of multistability are reviewed in \cite{PisarchikPR}. 

The dynamics of two unidirectionally coupled systems (master-slave configuration) of R\"ossler-like \cite{Pisarchik06, Pisarchik08a}, Duffing \cite{Zhu08} and R\"ossler-Lorenz \cite{Guan05} oscillators, as well as H\'enon maps \cite{Sausedo-Solorio11} have been studied, and some experiments along these lines have been carried out \cite{Pisarchik08b}. Bidirectionally coupled neuronal models \cite{Gu13} also display very rich synchronous dynamics. One of the most prominent features of all these examples is the intricate dependence of synchronization on the initial conditions, a distinct feature of multistable systems that is nowhere to be found in monostable systems. Phenomena such as anticipated intermittent phase synchronization, period-doubling synchronization, and intermittent switches between coexisting type-I and on-off intermittencies have been discovered \cite{Pisarchik08a, Pisarchik08b}. On the other hand, even though the existence and stability of multistable synchronous solutions in locally coupled Kuramoto models have been studied \cite{Wiley2006,Tilles2011,Tilles2013}, to our knowledge, the issue of under what conditions synchronization of more than two coupled generic (and possibly chaotic) multistable oscillators is guaranteed has not been addressed yet. Furthermore, synchronization of multistable systems in the presence of intermittency still remains an unexplored problem.

In this paper, we propose a methodology for studying the synchronization of multistable oscillators, which we illustrate with the example of a bistable system which has the great advantage of being experimentally   implemented in electronic circuits. 
First, we demonstrate the high complexity of the basins of attraction of coexisting states in a solitary bistable oscillator, and the increasing complexity when two of such oscillators interact with each other giving rise to intermittency. Second, we investigate the influence of both the initial conditions and the coupling strength on the synchronization of two bidirectionally coupled bistable systems (with diffusive coupling) in different coexisting synchronous states, including the existence of intermittency. Then, we discuss the Master Stability Function (MSF) \cite{PCmaster} approach to the study of the stability of a synchronization manifold of $N$ coupled multistable systems. Specifically, 
 we obtain the MSF for different coexisting chaotic attractors in a dynamical system separately,  and then we evaluate how the modification of the coupling parameter allows the system to leave/enter a particular synchronization regime associated with a particular attractor without loss of synchrony in the whole network, even in the presence of intermittency. 
Finally, we check the robustness of our theoretical predictions with electronic circuits to show the validity of our results for real systems where a certain parameter mismatch always exists.
\vskip-5cm
\section{Characterization of the system in terms of attractors and their basins of attraction}

The main aim of our work is to develop a methodology
that adequately predicts synchronizability in ensembles of bidirectionally coupled multistable systems. With this aim, we choose the piecewise linear  R\"ossler oscillator as a paradigmatic example of a bistable system with two coexisting chaotic attractors \cite{Pisarchik08b}. When an unidirectional coupling is introduced, the coupled R\"ossler-like system exhibits very rich dynamics, including such phenomena as intermittency, frequency-shifting and frequency-locking \cite{Pisarchik06}. Nevertheless, little is known about synchronization scenarios for ensembles of bidirectionally coupled multistable systems, despite the fact that bidirectional coupling itself can lead to the emergence of multiple attractors \cite{yanchuk2001}.

Specifically, the equations describing the dynamics of the R\"{o}ssler-like oscillators are \cite{Pisarchik06}
\begin{equation}
\begin{array}{l}
\dot{x}=-\alpha_1 (x+\beta y+\Gamma z), \\
\dot{y}=-\alpha_2 (-\gamma x-[1-\delta] y), \\
\dot{z}=-\alpha_3 (-g(x)+z),
\label{Rossler1}
\end{array}
\end{equation}
with
\begin{equation}
g(x)=\left\{
\begin{array}{cc}
0 & x\leq 3, \\
\mu \left( x-3\right)  & x>3,%
\end{array}%
\right.  \label{geq}
\end{equation}%
where $x$, $y$, and $z$ are the state variables. The piecewise linear function $g(x)$ introduces the nonlinearity in the system that leads to a chaotic behavior. The parameter values are $\alpha_1=500$, $\alpha_2=200$, $\alpha_3=10000$, $\beta=10$, $\Gamma=20$, $\gamma=50$, $\delta=15.625$ and $\mu =15$. For this parameter choice, the system is known to be a bistable chaotic system (i.e. the phase portrait displays two different chaotic attractors), as previously reported in \cite{Pisarchik08b}. Unlike most previously studied multistable systems (see, e.g. \cite{Guan05}), this system exhibits multistability in the  autonomous evolution, without the need for chaotic driving. Moreover, this system can be implemented in electronic circuits to experimentally assess the validity of the theoretical predictions.

\begin{figure}[!t]
\includegraphics[scale=0.29]{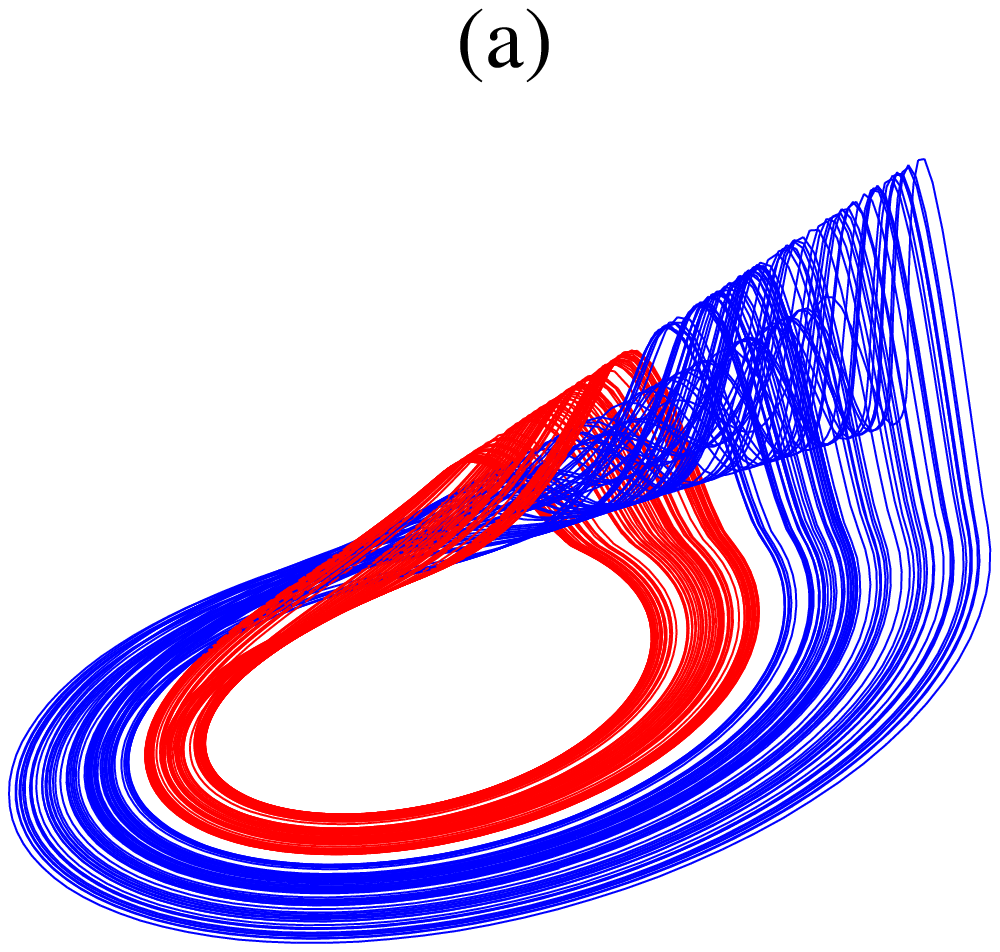}
\includegraphics[scale=0.29]{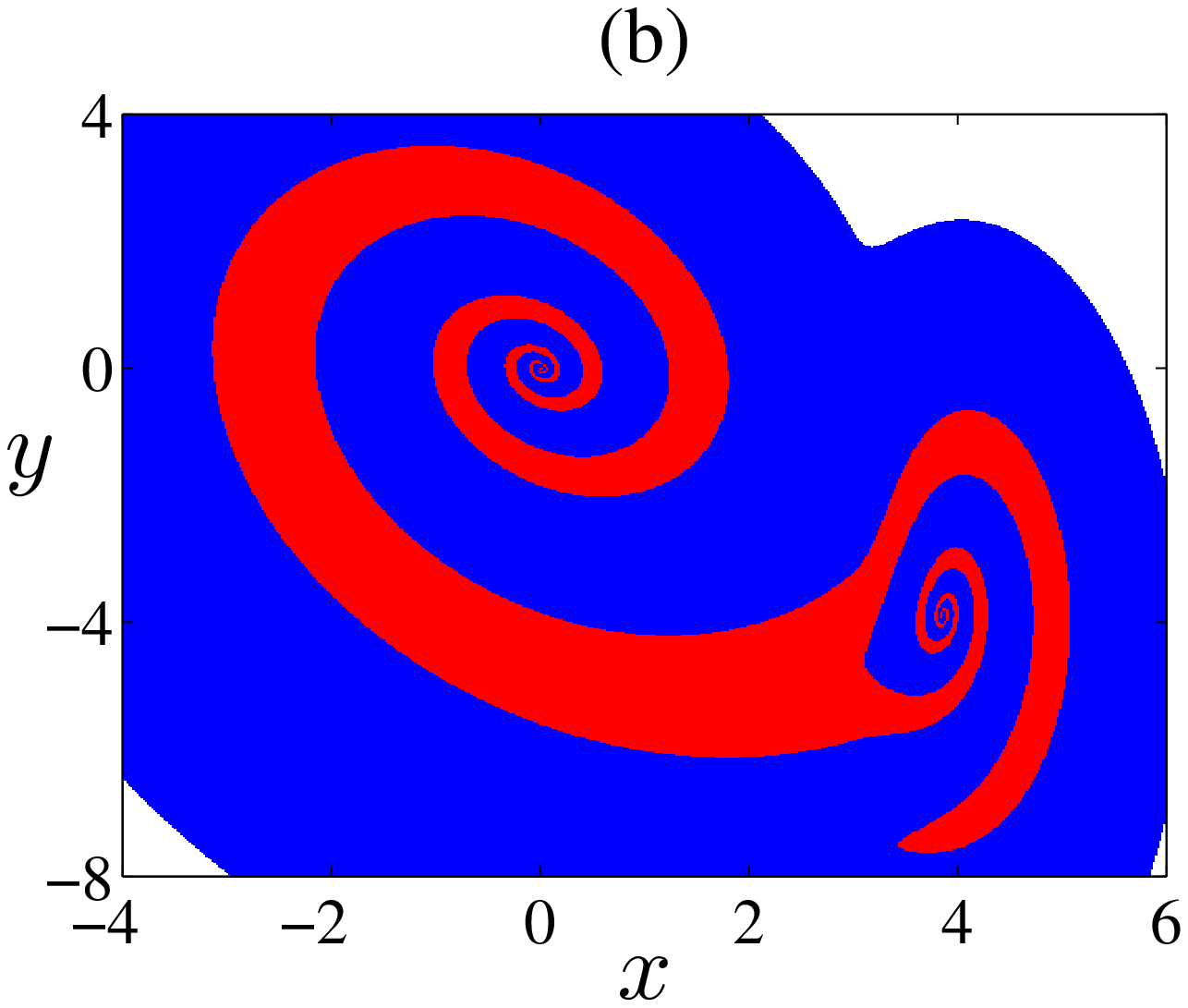}\\
\includegraphics[scale=0.29]{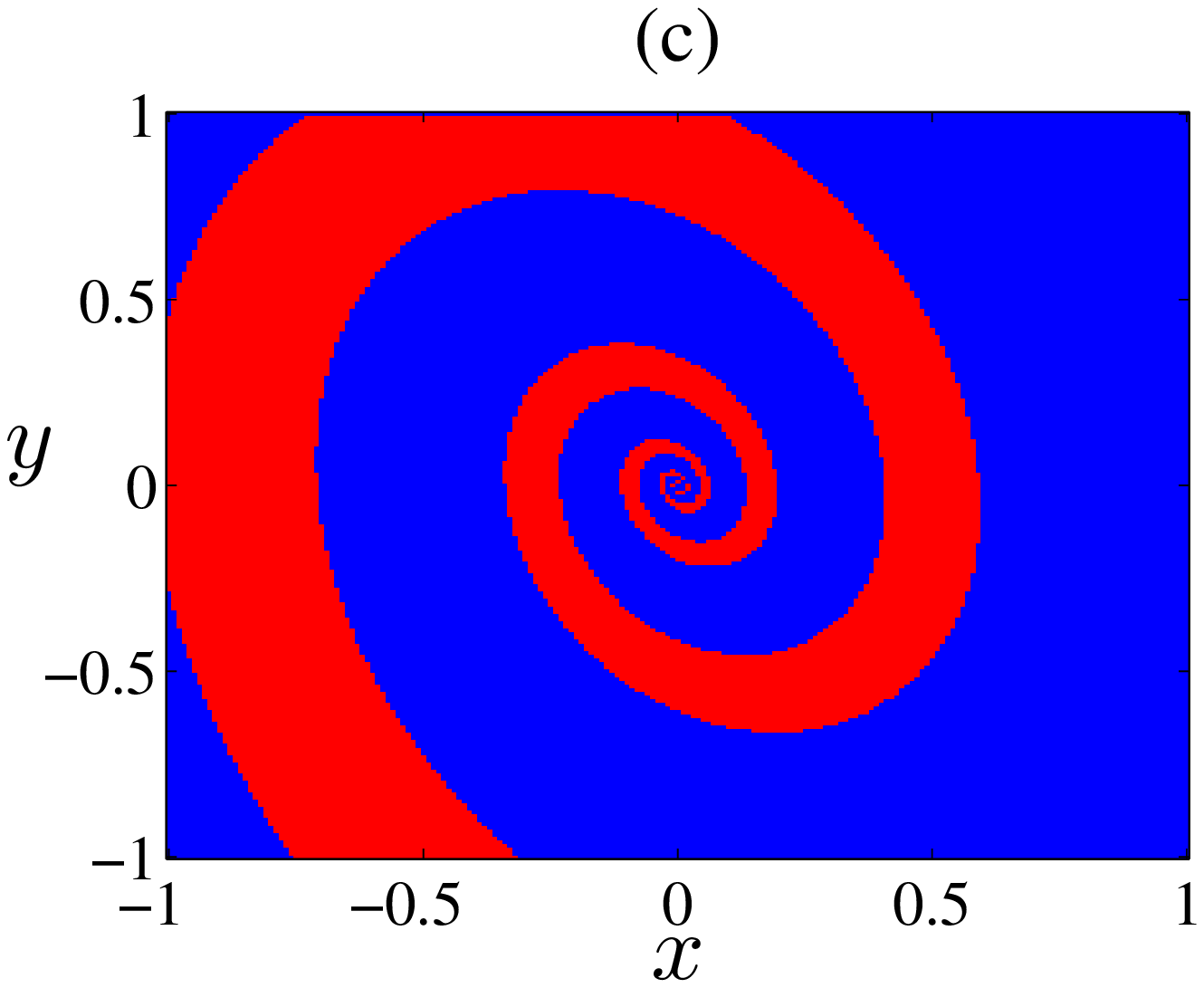}
\includegraphics[scale=0.29]{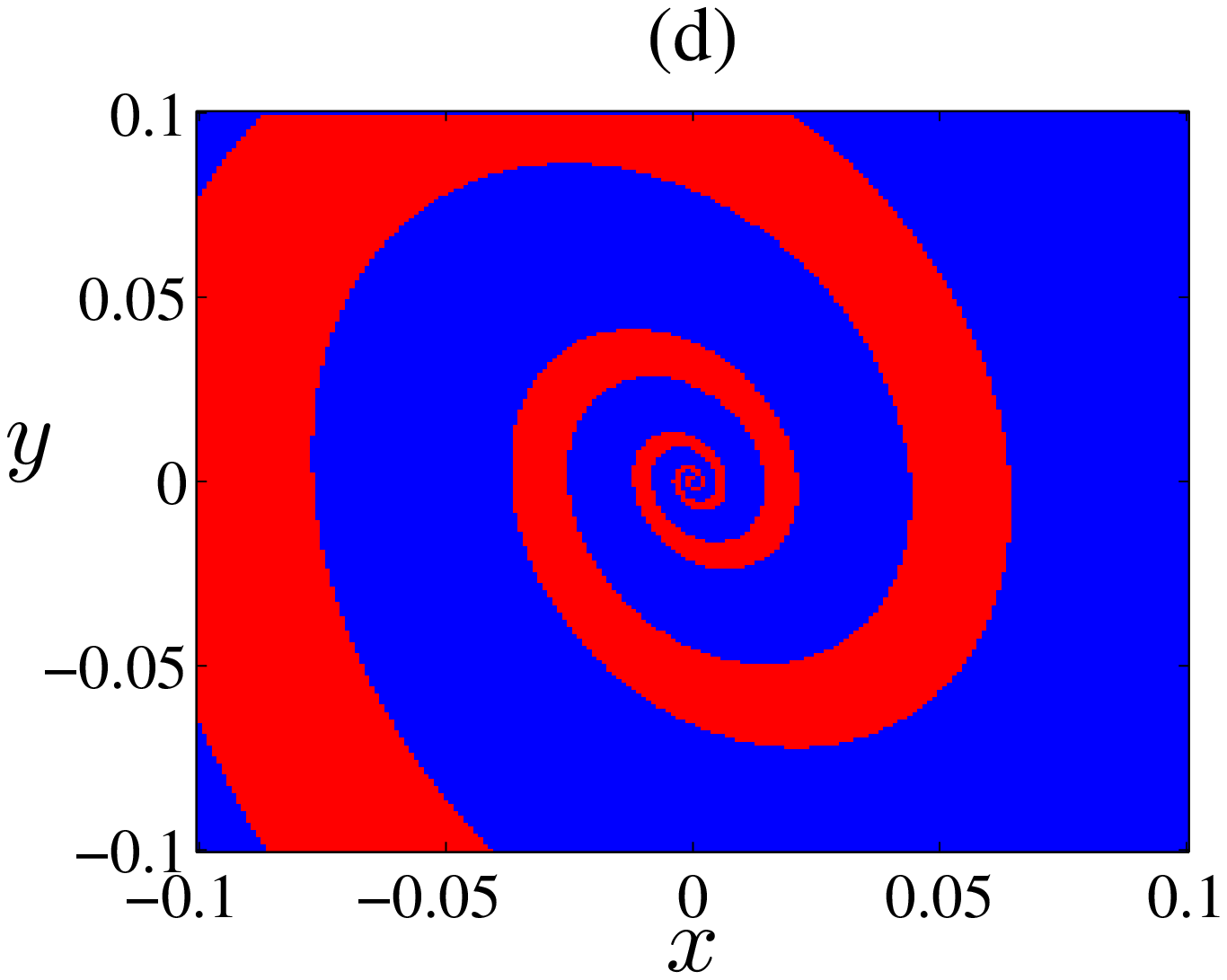}
\caption{\label{fig1} (Color online) {\bf Coexisting attractors and their basins of attraction in the R\"ossler-like oscillator (Eq. \ref{Rossler1}).} (a) Large ($L$, blue (dark gray)) and small ($S$, red (gray))} attractors.  (b) Basins of attraction for $L$ (blue) and $S$ (red) in the $z=0$ plane. Initial conditions leading to unstable trajectories appear in white. (c) Basins of attraction for $L$ and $S$, $[-1,1]\times[-1,1]$ square in the $z=0$ plane. (d)  Basins of attraction for $L$ and $S$, $[-0.1,0.1]\times[-0.1,0.1]$ square in the $z=0$ plane.
\end{figure}

From any arbitrary initial condition within a bounded region in phase space the system rapidly converges to one of the two chaotic attractors shown in Fig. \ref{fig1} (a). We denote the larger attractor by $L$ (Fig. \ref{fig1} (a), blue (dark gray)) and the smaller one by $S$ (Fig. \ref{fig1} (a), red (gray)). The basins of attraction of $L$ (blue (dark gray)) and $S$ (red (gray)) are shown in Fig. \ref{fig1} (b) for initial conditions ${\bf x}(0) \equiv (x(0),y(0),z(0))$ such that $x(0) \in [-4,6]$, $y(0) \in [-8,4]$ and $z(0) = 0$. The basin of attraction of $L$ is seen to be much larger than the basin of $S$. Two spirals are clearly visible, where initial conditions leading to one or the other attractor seem to be intertwined. Each of these spirals has a fixed point of the system as its focus, as has been previously reported in \cite{Pisarchik08a}. In Fig. \ref{fig1} (c) we focus on $x(0) \in [-1,1]$, $y(0) \in [-1,1]$ to better appreciate the details of the spiral that has its center in the origin. Indeed, the mixing of initial conditions close to the center of the spiral seems to be present at arbitrarily low space scales, as the zoom around $x(0) \in [-0.1,0.1]$, $y(0) \in [-0.1,0.1]$ in Fig. \ref{fig1} (d) shows. We have checked that this structure is preserved for another 4 orders of magnitude, with no end in sight for even lower space scales. Indeed, this is not altogether surprising, as basins that are interwoven in a complicated fashion and fractal basins boundaries feature prominently in the phase portraits of many multistable systems (see \cite{Aguirre09} for review of fractal basin boundaries and fractal sets in nonlinear dynamics in general). Although  the precise characterization of these boundaries is outside the scope of this paper, we would like to stress how difficult it is to control the asymptotic dynamics of just one single oscillator under the presence of noise or uncertainties for initial conditions starting in certain regions of the phase space.

\section{Two bidirectionally coupled systems: coexisting regimes}

To start our study on synchronization of bidirectionally coupled multistable systems, we first consider the simple case of two coupled R\"{o}ssler-like oscillators. This particular system has been thoroughly analyzed for the case of a master-slave configuration \cite{Pisarchik06,Pisarchik08a,Pisarchik08b}. Here, we investigate, for the first time to our knowledge, bidirectionally coupled multistable chaotic systems, and also for the first time the coupling is diffusive. The coupling is introduced through the $x$ variable with coupling strength $\sigma$ so that the equations of the motion become
\begin{equation}
\begin{array}{l}
\dot{x}_{1,2}=-\alpha_1 (x_{1,2}+\beta y_{1,2}+\Gamma z_{1,2}-\sigma\psi\left(x_{2,1}-x_{1,2}\right)), \\
\dot{y}_{1,2}=-\alpha_2 (-\gamma x_{1,2}-[1-\delta] y_{1,2}),\\
\dot{z}_{1,2}=-\alpha_3 (-g(x_{1,2})+z_{1,2}),
\label{Rossler2}
\end{array}
\end{equation}
where $\psi = 20$ and $g(x)$ is given by Eq. (\ref{geq}). Our numerical simulations show the existence of four possible asymptotic regimes: a) both systems end up in an attractor indistinguishable from $L$ (from now on, $LL$), b) both systems end up in an attractor indistinguishable from $S$ ($SS$), c) one system asymptotes to $L$ and the other to $S$ ($SL$), d) the systems switch intermittently back and forth between $L$ and $S$ in an irregular way ($I$) (the intermittent behavior of these systems in master-slave configurations is described in \cite{Pisarchik08b}).

\begin{figure}[!b]
\includegraphics[scale=0.29]{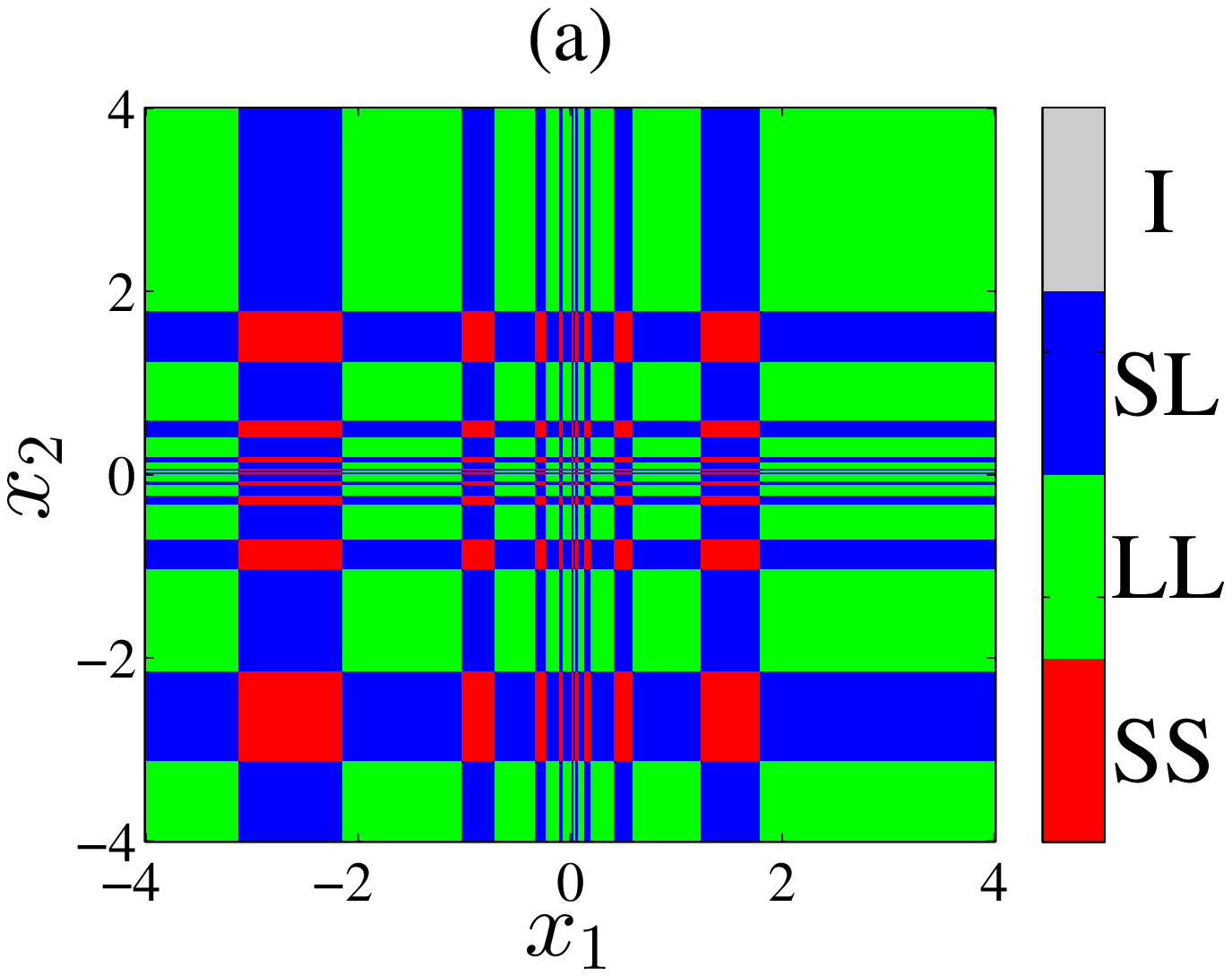}
\includegraphics[scale=0.29]{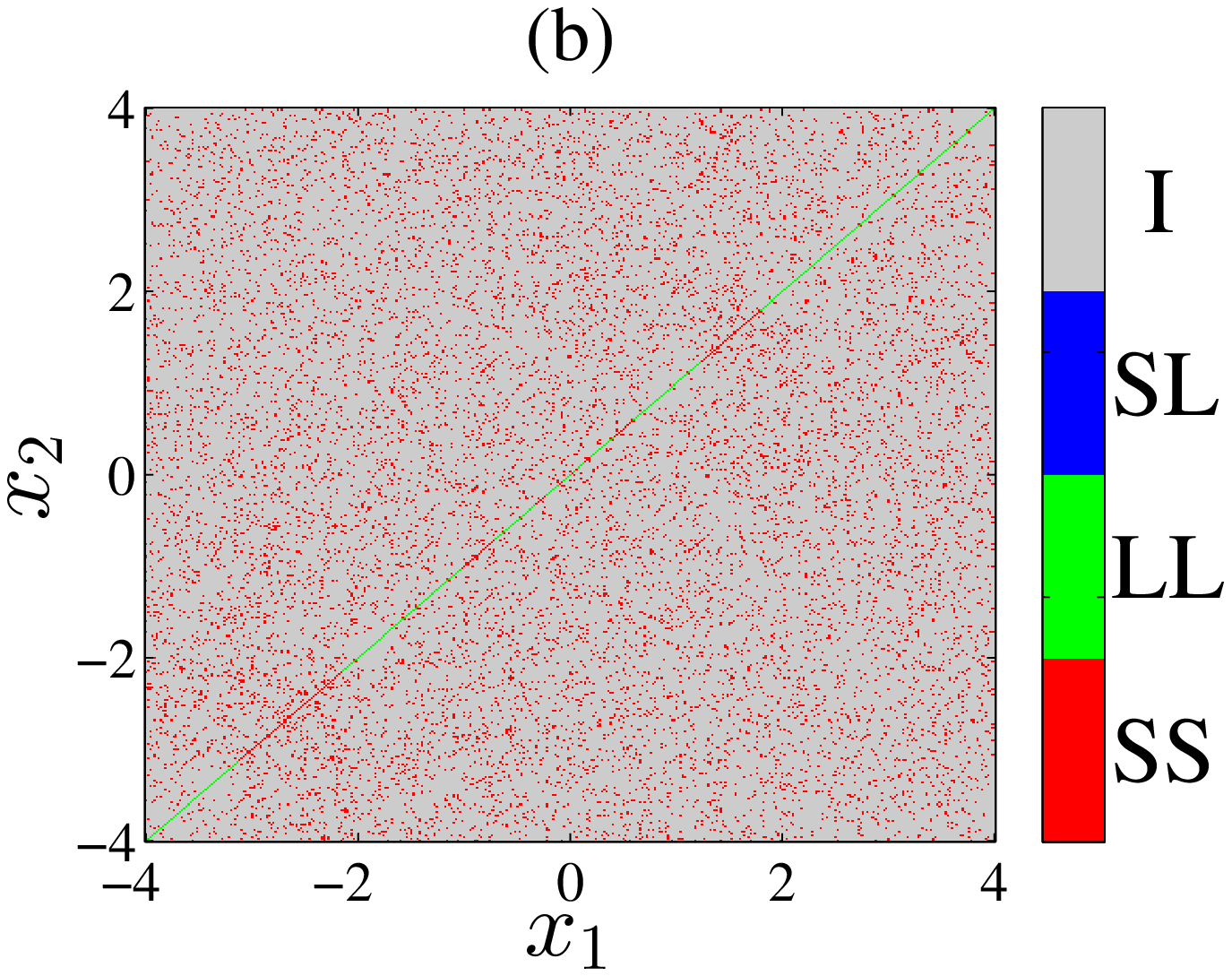}\\
\includegraphics[scale=0.29]{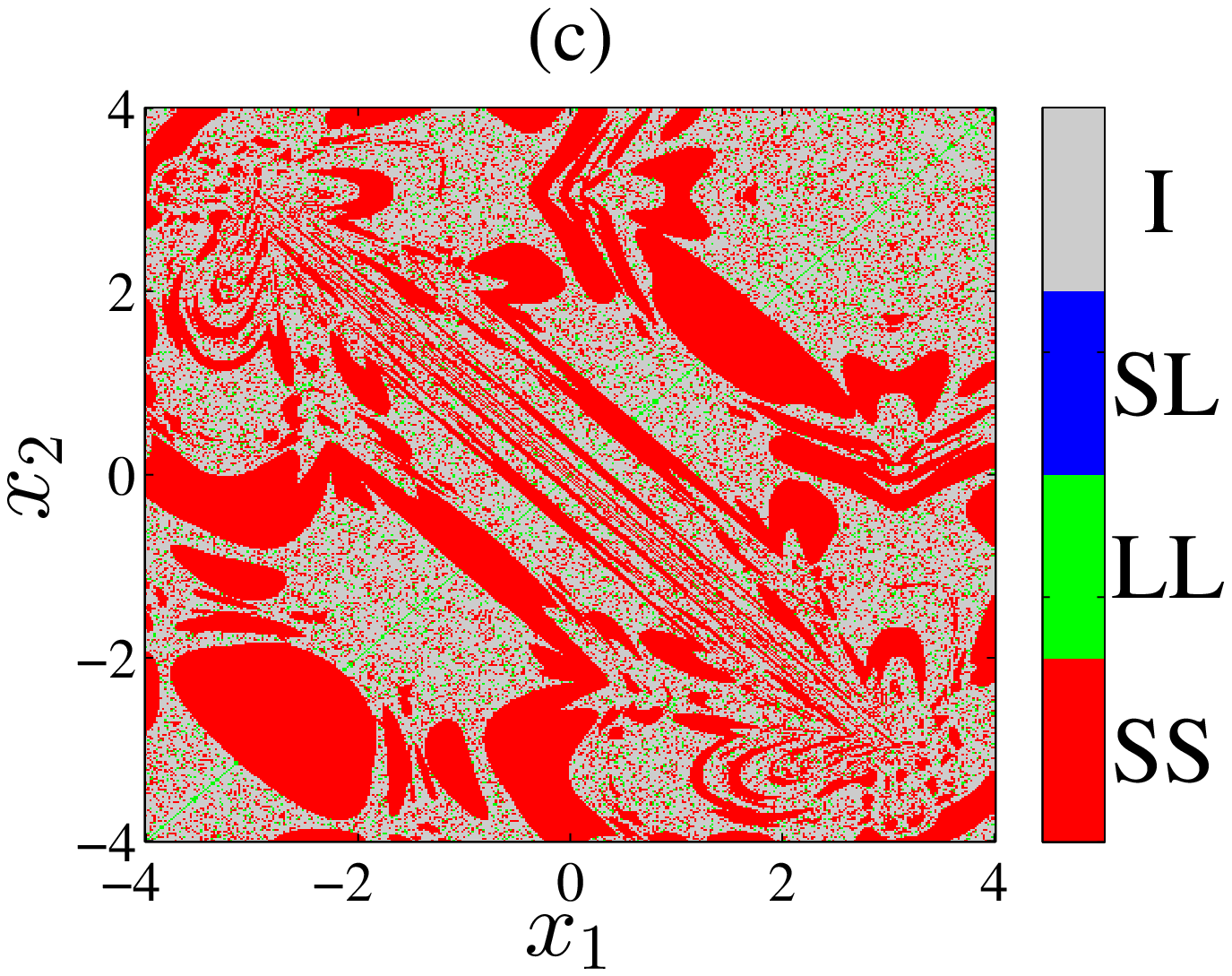}
\includegraphics[scale=0.29]{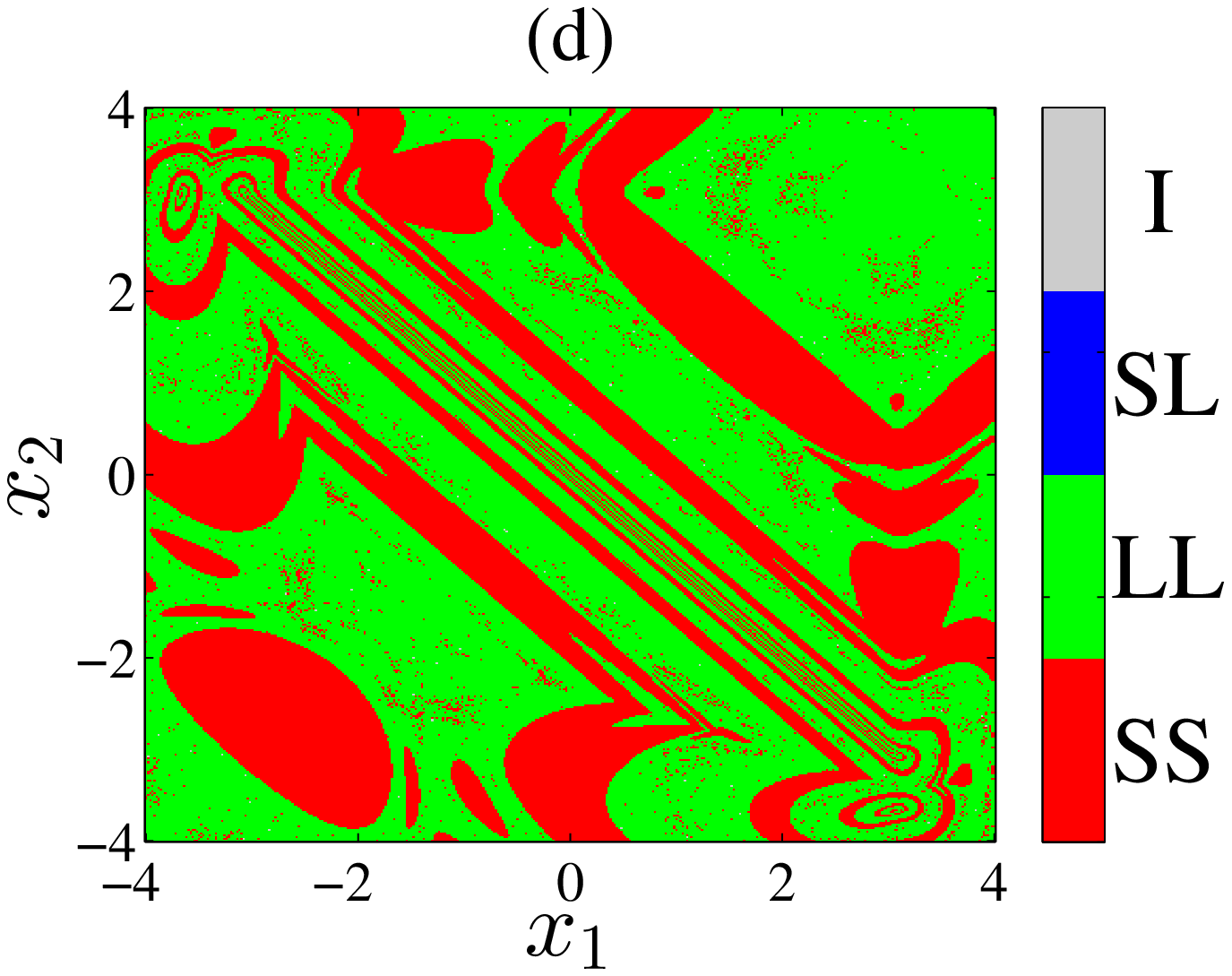}
\caption{\label{fig2} (Color online) {\bf Basins of attraction of the system comprising two bidirectionally coupled R\"ossler-like oscillators for different coupling strengths $\sigma$}. The coupling strengths are (a) $\sigma = 0$ (uncoupled systems), (b) $\sigma = 0.06$, (c) $\sigma = 0.14$, and (d) $\sigma = 0.20$.}
\end{figure}

As the coupling strength $\sigma$ is increased from $0$ to $0.2$, all these cases appear, disappear and mix in a very complicated manner, depending on the initial conditions. In Fig. \ref{fig2} we show the basins of attraction for these asymptotic regimes that result from fixing $y_1(0) = y_2(0) = 0$ and $z_1(0) = z_2(0) = 0$, and exploring a finely discretized grid for $x_{1,2}(0) \in [-4,4]$ for four representative coupling strengths, $\sigma = 0$, $0.05$, $0.14$, and $0.2$. For moderately larger $\sigma$ the basins remain relatively stable, while the phase space projections of the full attractor on the subspace corresponding to each of the two subsystems display only very slight deformations with respect to the original $L$ and $S$ seen for $\sigma = 0$.

\begin{figure}[!t]
\includegraphics[scale=0.29]{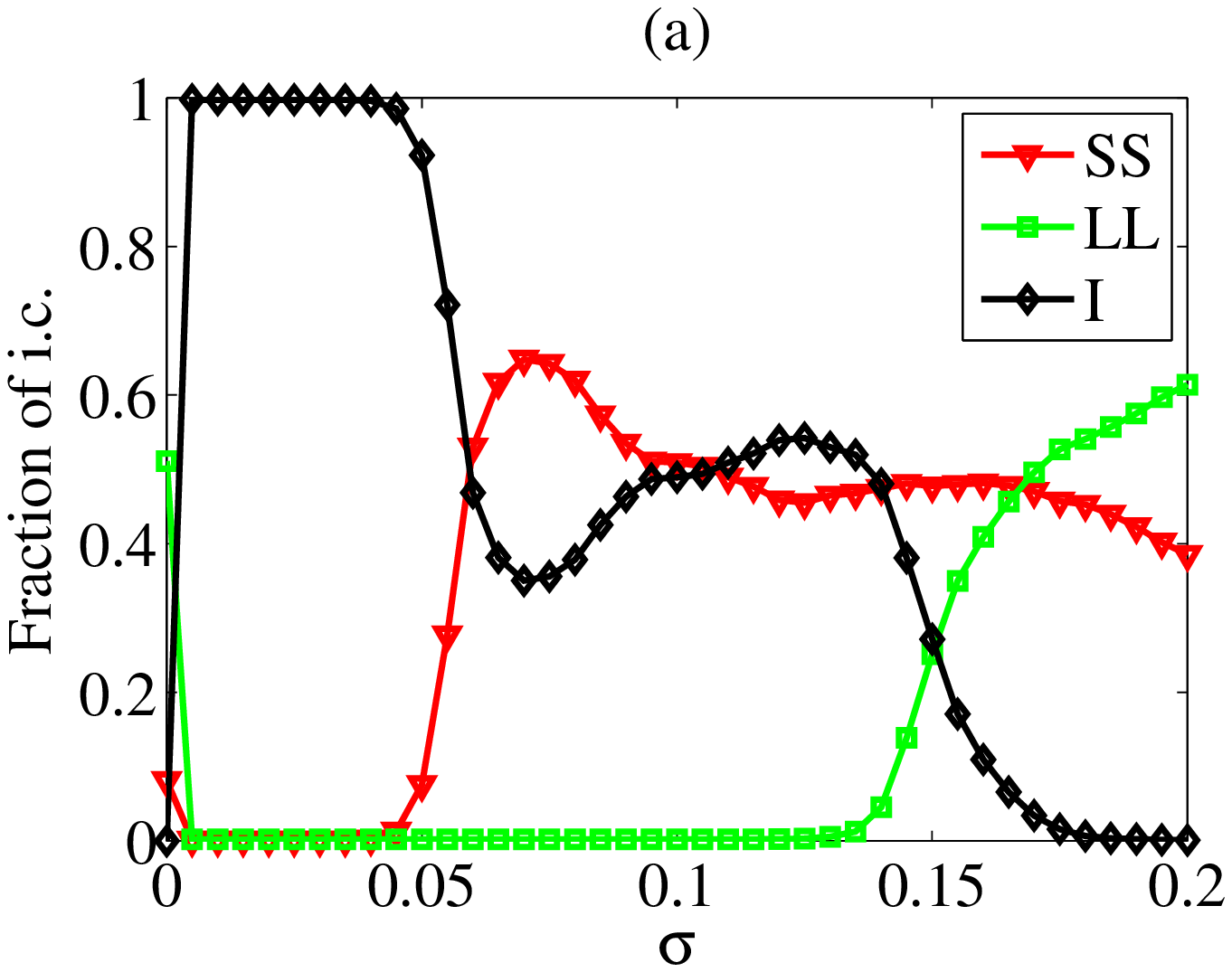}
\includegraphics[scale=0.29]{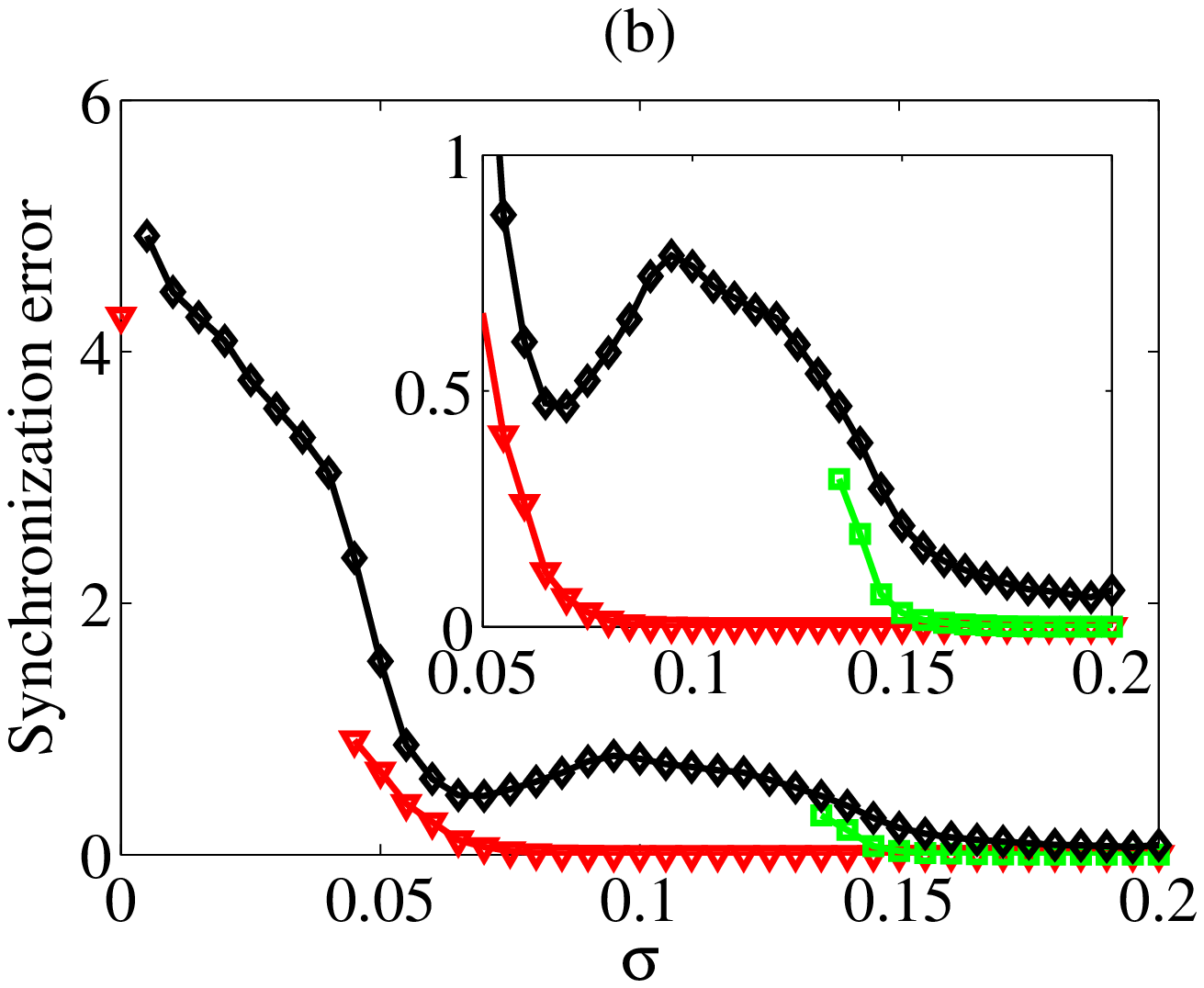}\\
\includegraphics[scale=0.29]{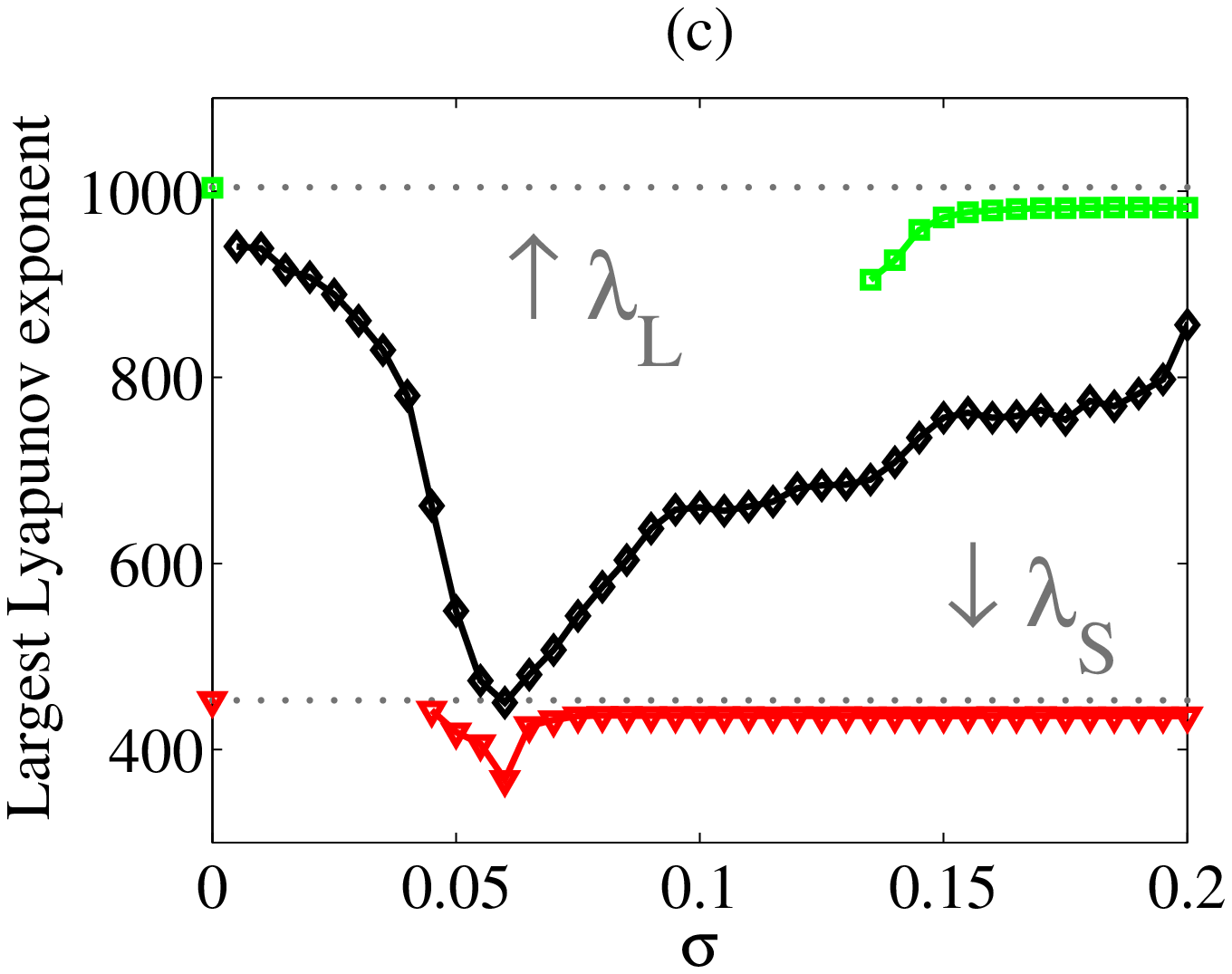}
\includegraphics[scale=0.29]{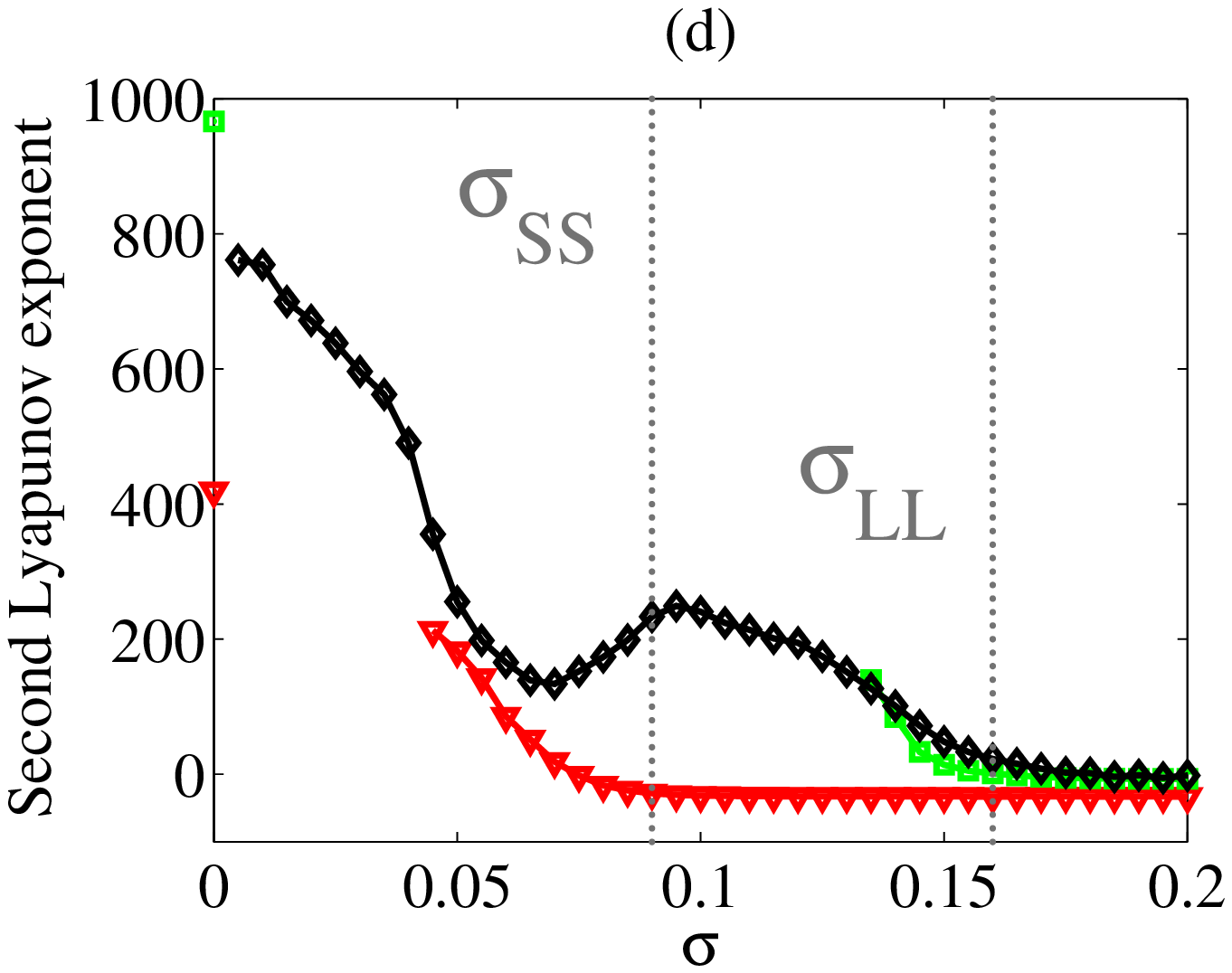}
\caption{\label{fig3} (Color online) {\bf Relative size of the basins for each attractor, synchronization error, largest and second largest Lyapunov exponents of the full system as a function of $\sigma$.} (a) Relative size of the basins of attraction of the full system, (b) synchronization error (inset: zoomed part of the plot), (c) largest Lyapunov exponent, (d) second largest Lyapunov exponent.}
\end{figure}

For better visualization of the prevalence of the different asymptotic regimes for different values of $\sigma$, Fig. \ref{fig3} (a) shows the fraction of the initial conditions considered, which eventually lead to every possible asymptotic regime, as a function of the coupling strength. As expected from Fig. \ref{fig2}, Fig. \ref{fig3} (a) shows that for small $\sigma$ most of the initial conditions lead to the intermittent behavior (Fig. \ref{fig2} (b)), whereas stronger coupling leads first to the emergence of $SS$ (Fig. \ref{fig2} (c)) and later to the emergence of $LL$ together with a vanishing of intermittent behavior $I$. For large $\sigma$, only $LL$ and $SS$ are observed (Fig. \ref{fig2} (d)). Interestingly, the $SL$ case is never observed for nonzero $\sigma$, so that we do not consider it anymore.

\section{Two bidirectionally coupled systems: synchronization of different regimes}

Having seen the different possible asymptotic regimes and their basins of attraction, we move on to study synchronization of two bidirectionally coupled systems. Instead of focusing on individual trajectories, we average the results across all initial conditions leading to one of the considered cases ($LL$, $SS$ or $I$). 
First, we calculate the synchronization error (Fig. \ref{fig3} (b)) for each of the different asymptotic regimes. This way, we observe how synchronization is first achieved for the $SS$, red (gray) triangles, case around $\sigma \simeq 0.075$, and for the $LL$, black (dark gray) diamonds, case only for $\sigma \simeq 0.165$. As for the $I$, green (light gray) squares, case, the synchronization depends on the coupling in a more complex, non-monotonic manner, which in some way reflects the synchronization patterns of $SS$ and $LL$. As the system becomes more and more synchronized and therefore the influence of one subsystem on the other becomes weaker, $I$ fades away.
The transition to synchronization
in this regime is consistent with the observation made
in \cite{Zhu08}, as it passes through two stages: first, the two oscillators
move to the same attractor due to the increasing coupling,
and second, as the coupling further increases, they gradually
synchronize (just as monostable coupled oscillators do).

A different view on the emergence of synchronization is provided by the analysis of the Lyapunov spectrum. Figure \ref{fig3} (c) shows the largest Lyapunov exponent as a function of $\sigma$ for different asymptotic regimes. The dotted lines show the largest Lyapunov exponents corresponding to the attractors $L$ and $S$ for the solitary (uncoupled) system. In the $I$ regime, the value of the Lyapunov exponent is clearly correlated with the presence of $LL$ and $SS$ windows, a phenomenon that will be elucidated below. However, from the synchronization point of view, the most important information derived from the Lyapunov spectrum concerns the second largest Lyapunov exponent, which gives indirect information on the existence of weakest forms of synchronization, such as generalized synchronization (GS) (see, for instance, the discussion in  \cite{Gut13}). As seen in Fig. \ref{fig3}(d), two points, highlighted as $\sigma_{SS}$ and $\sigma_{LL}$, corresponding to the loss of the positive Lyapunov exponent in the system, mark the starting point of the existence of just one chaotic mode in the system, and therefore the onset of GS for $SS$ and $LL$ respectively. Also, $I$ seems to reach GS, but as we will see $I$ is a very sophisticated regime, with frequent jumps between time windows, where both oscillators are in the same attractor, and time windows,  where each oscillator is in a different attractor. Time averages across very long time windows (as those used in the calculation of the synchronization error or Lyapunov exponents) fail to capture this sort of details, so we presently conclude this section with a brief study of this regime to illustrate just how complex it can be.

\begin{figure}[!b]
\includegraphics[scale=0.29]{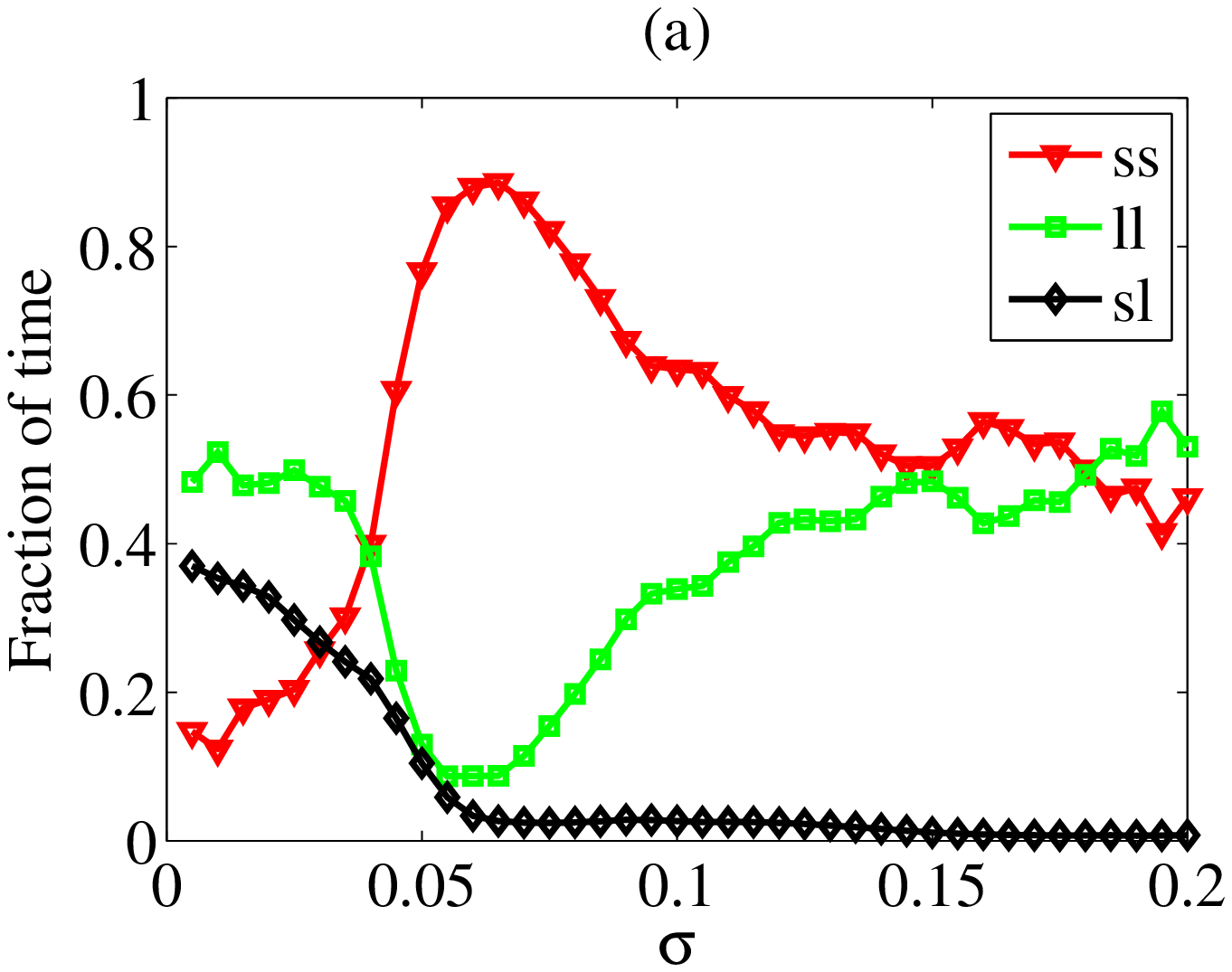}
\includegraphics[scale=0.29]{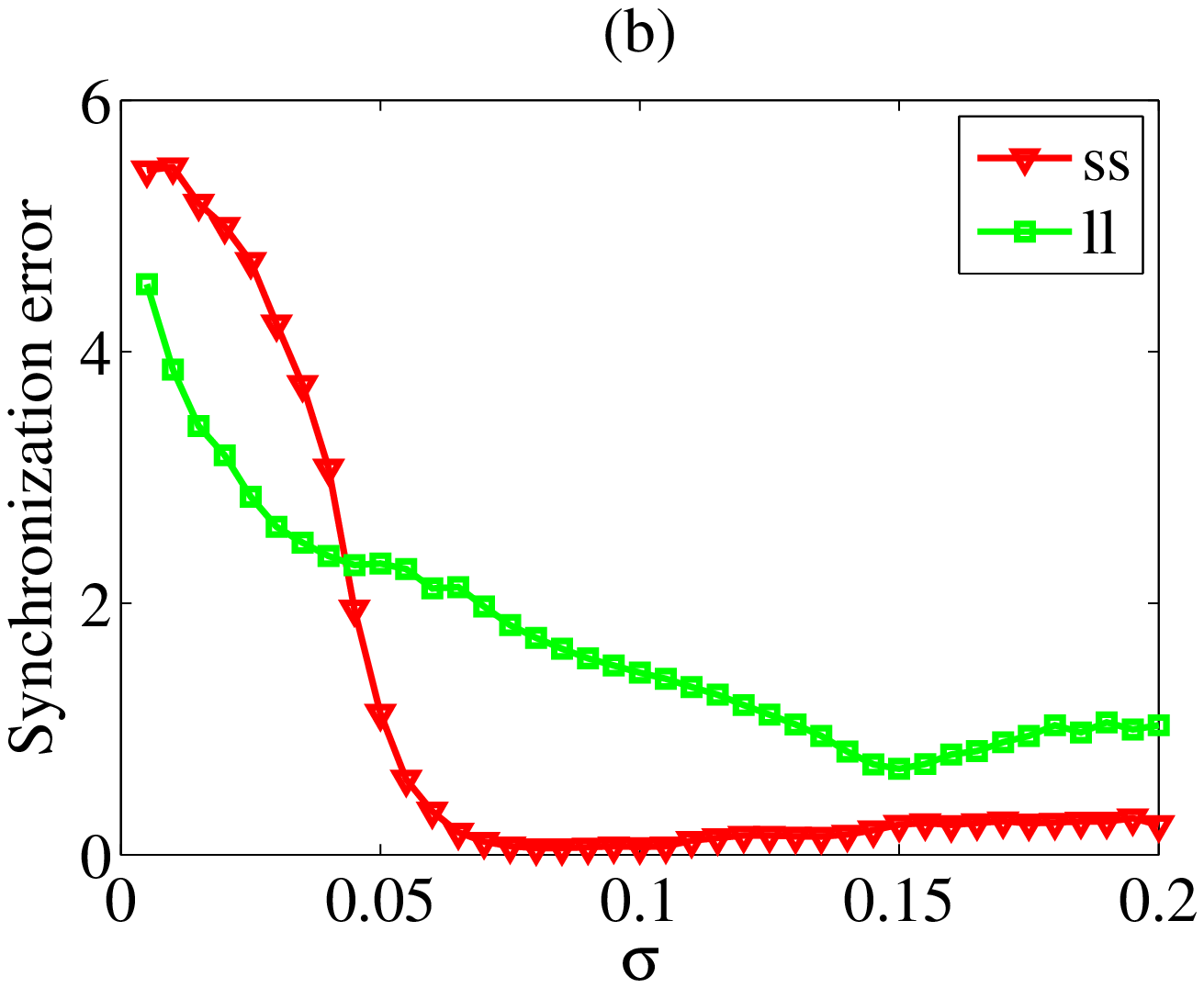}
\caption{\label{fig4} (Color online) {\bf Fraction of time and synchronization error in the $I$ regime for different time windows as a function of $\sigma$.} (a) Fraction of the time spent by the system in $ss$, $ll$, and $sl$  windows, (b) $I$ synchronization error restricted to $ll$ and $ss$ time windows.}
\end{figure}

We now focus on the dynamics of the coupled system in the $I$ regime and measure the fraction of time spent by both subsystems simultaneously in the $S$ state (denoted as $ss$), $L$ state ($ll$), and when each subsystem follows a different dynamics ($sl$). Figure \ref{fig4} (a) shows the fraction of time that the system being in the $I$ regime spends in $ss$, red (gray) triangles, $ll$, green (light gray) squares, and $sl$, black (dark gray) diamonds, time windows, which bears an obvious relation to the results shown in Fig. \ref{fig3} (c).  Another important issue is the lack of complete synchronization in $I$ even for very high $\sigma$. To examine this problem in more detail, we now compute the synchronization error separately for the $ss$ and $ll$ time windows. We do not consider the $sl$ case because by definition complete synchronization is not possible in these time windows. In Fig. \ref{fig4} (b) the synchronization error reveals that for $\sigma > 0.075$ the $I$ regime contains completely synchronized $ss$ time windows, while the $ll$ time windows show a clear lack of synchronization. It is difficult to ascertain whether the mild growth in the synchronization error of $ss$ for $\sigma > 0.10$ is indeed a loss of synchronization or has to do with the difficulties associated with the classification of time windows into $ss$, $ll$, and $sl$ \cite{detectcriterion}.  In the next section we will present a straightforward way to understand the conditions under which synchronization is possible in an ensemble of multistable dynamical systems even in the presence of intermittent regimes such as the one we just described. 
 
\section{Master Stability Function of multistable dynamical systems}

Up to now, we have seen how complex the basins of attraction of the bistable oscillators are, and how they become much more complex as soon as two oscillators interact, with the development of new phenomena such as intermittency, that makes synchronization predictions very difficult and extremely dependent on initial conditions. Whenever multistable oscillators are coupled in a large ensemble, we expect the dynamics to become even more complex, and the problems associated with the study and control of synchronization become even more serious. In order to measure the synchronization stability of ensembles of coupled multistable oscillators, we resort to the Master Stability Function (MSF) approach of Pecora and Carroll \cite{PCmaster}. The MSF curve is given by the maximum of the Lyapunov exponents transverse to the synchronization manifold. In the past 15 years, the MSF approach has found important applications in the study of networks of monostable systems (see, e.g. \cite{CMSF}), however, as far as we know, it has never been extended to the case of systems with coexisting attractors.

\begin{figure}[!t]
\includegraphics[scale=0.30]{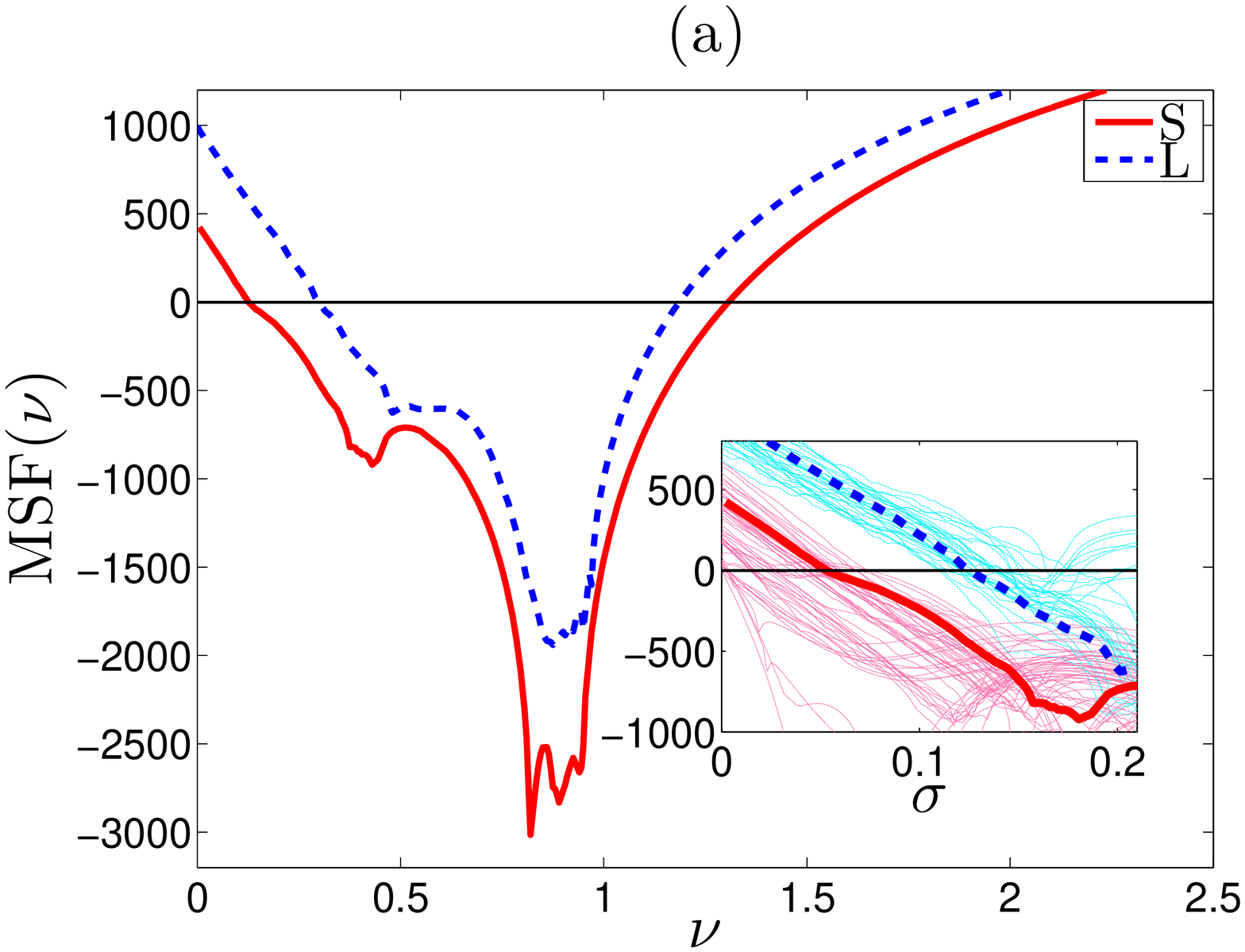}
\includegraphics[scale=0.30]{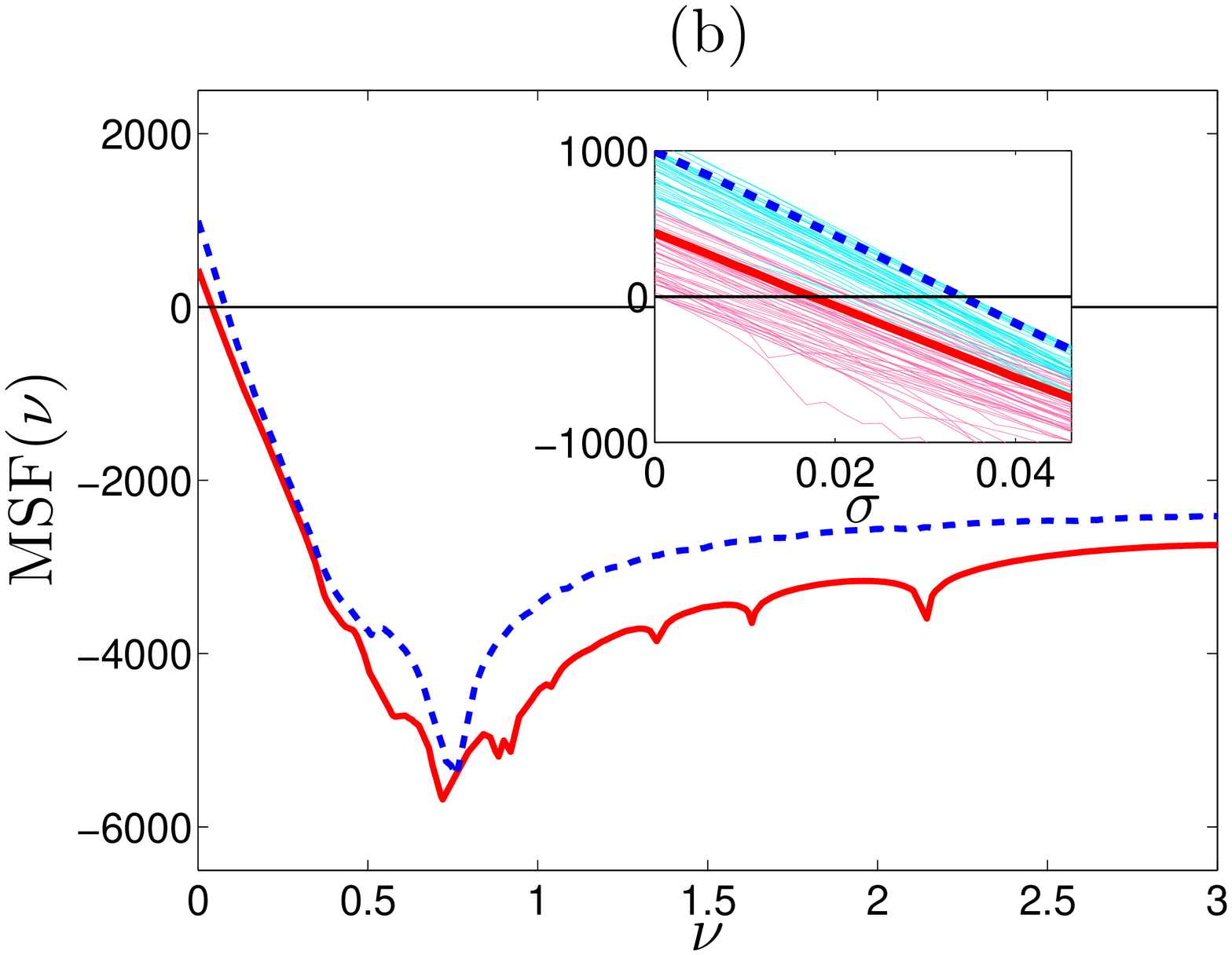}
\caption{\label{fig5} (Color online) {\bf MSFs for the coexisting attractors for two types of coupling, and variability with respect to uncertainties in the parameters.} (a) MSF of $S$, red (gray) continuous line, and $L$, blue (dark gray) dashed line, for oscillators coupled through the $x$ variable (class III system); inset: variability of the first zero crossing as the parameters are affected by uncertainties (see Section {\it Experimental Implementation}  for a detailed explanation). (b) MSF of $S$ (red continuous line) and $L$ (blue dashed line) for oscillators coupled through the $y$ variable (class II system); inset: variability of the zero crossing as the parameters are affected by uncertainties.}
\end{figure}

While the MSF describes the linear stability of the synchronous motion for a given attractor dynamics, in the presence of multistability one should rather look at the MSF of each attractor separately. In Fig. \ref{fig5} (a) we show the MSFs corresponding to both $S$, red (gray) continuous line, and $L$, blue (dark gray) dashed line, for the R\"ossler-like system with diffusive coupling through the $x$ variable as in Eq. (\ref{Rossler2}). The independent variable $\nu$ is a parameter that implicitly takes into account infinitely many network topologies and coupling strengths, as it stands for the product of a given coupling strength $\sigma$ and a graph Laplacian eigenvalue (for each eigenmode transversal to the synchronization manifold there is one such eigenvalue, and there are $N-1$ of them for a fully connected network of size $N$). According to the classification proposed in \cite{CMSF}, this system belongs to the class III systems, as the region of stable synchronization is bounded between two MSF zeros. Interestingly, Fig. \ref{fig5} also shows that the stability region of $L$ is contained in that of $S$. The interplay between the two stability regions ($S$ and $L$) given by the MSFs explains how the synchronized state is maintained (or lost) in one of the three possible scenarios $SS$, $LL$, and $I$. There are therefore three possibilities for a given $\sigma$ and a given eigenmode of the topology (i.e. for a given $\nu \equiv \sigma \lambda$, where $\lambda$ is the specific eigenvalue of the network's Laplacian matrix that corresponds to the considered eigenmode): a) neither $S$ nor $L$ can be synchronized (i.e., the value of the MSFs of both $S$ and $L$ are positive for such $\nu$), b) $S$ can be synchronized, but not $L$ (i.e., the MSF is negative for $S$, but positive for $L$), and c) both $S$ and $L$ can be synchronized (both values of the MSFs are negative). While b) only guarantees the synchronization stability in the $SS$ regime, c) is tantamount to saying that {\it synchronization is stable no matter how complex the dynamics may be, even in the presence of intermittency}. Therefore, if the product of each of the eigenmodes for a given topology and $\sigma$ is within the region described by c), synchronization is stable. Analogous arguments can be used in systems with an arbitrarily large number of attractors, provided the stability regions of the different attractors are not disjoint.

Figure \ref{fig5} (b) shows the MSFs corresponding to the two coexisting attractors when the oscillators are coupled through the $y$ variable. In this case the system belongs to class II, i.e. the stability region starts at a given $\nu$ and then extends indefinitely to the right. Again, for this particular system the stability region of $L$ is contained in that of $S$. Choosing $\sigma$ high enough (or changing suitably the topology for a given $\sigma$) in such a way that all eigenmodes are in the region lying to the right of the MSF zero for $L$ guarantees the stability of synchronous dynamics on any attractor, even in the presence of intermittency. Below we will dwell upon the contents of the insets in Figs. \ref{fig5} (a) and (b).

\section{Experimental implementation}

The MSF approach determines the stability of the synchronization manifold but does not give information about how large the {\it stability basin} of the system is. Nevertheless, in real systems, the size of the stability basin of the synchronized manifold plays a fundamental role in the observation of synchronization, specially when external perturbations are applied (see \cite{menck2013} for a detailed description of how to evaluate a stability basin). In this section, we compare our theoretical predictions given by applying the MSF approach to multistable systems with experimental results in order illustrate the robustness under the presence of noise and parameters mismatches, and verify their validity even when the noise and the coupling of non-identical systems bring the system quite far away from the synchronization manifold. 

\begin{figure}[!b]
\includegraphics[scale=0.14]{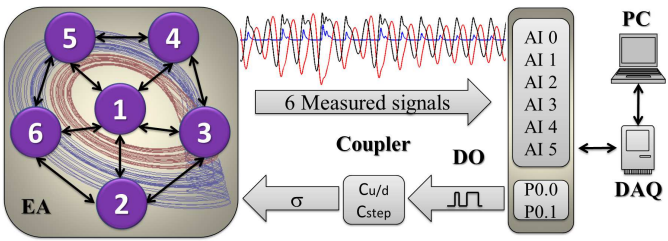}
\caption{ \label{fig6} (Color online) {\bf Experimental setup.} On the left, schematic representation of the coupling topology of the 6-circuit network.  The coupling is adjusted by a digital potentiometers X9C104, whose parameters C$_{u/d}$ (Up/Down resistance) and C$_{step}$ (increment of the resistance at each step) are controlled by a digital signal coming from a DAQ Card, P0.0-P0.1 respectively. The outputs of the circuit are sent to a set of voltage followers that act as a buffer and, then, sent to the analog ports (AI 0 ; AI 1; ... ; AI 5) of the same DAQ Card. The whole experiment is controlled by a PC using a Labview Software.}
\end{figure}

The experimental design is based on a 6-node network of piecewise R\"{o}ssler-like electronic circuits coupled according to the topology shown in Fig. \ref{fig6}, i.e. following a {\it spiderweb network} topology with a central node connected to all other nodes, and each of the 5 peripheral nodes connected to their 2 spatial neighbors. Other network configurations, with an arbitrary number of oscillators, are also possible, the only limitation being the number of available electronic components. The circuits are the experimental implementation of the dynamical system given by Eq. (\ref{Rossler1})  (see \cite{Pisarchik08b} for a detailed description of the experimental realization of the R\"ossler-like oscillator, and \cite{Exp12,Aguirre14} for previous realizations in network configurations). As in the evaluation of the MSFs illustrated in Fig. \ref{fig5}, we consider the system coupled through the $x$ and then through the $y$ variable, the  cases being paradigmatic examples of class III and class II systems, respectively.

The experimental setup shown in Fig. \ref{fig6} consists of an electronic array (EA), a multifunction data card (DAQ), and a personal computer (PC). The EA comprises 6 R\"ossler-like electronic circuits forming the spiderweb network with one central node and 5 peripheral nodes. Each node has an individual electronic coupler controlled by a digital potentiometer (XDCP), which is adjusted by a digital output signal (DO) coming from ports P0.0 and P0.1. Port P0.0 is used to set the value of the coupling resistance (adequately scaled to correspond to the values of $\sigma$) and P0.1 increases or decreases the value of the resistance through  a voltage divisor (the resolution allowing for 100 discretized steps). The full experimental process is controlled with a virtual interface developed in Labview 8.5, that can be considered as a state machine. The experimental procedure is realized as follows. First, $\sigma$ is set to zero, after a waiting time of 500 ms (roughly corresponding to $600$ cycles of the autonomous systems), the output signals from the 6 circuits are acquired by the analog ports (AI 0; AI 1; ...; AI 5). Once the dynamics of the whole ensemble is recorded, the value of $\sigma$ is increased by one step, and the signals are again stored in the PC for further analysis. This process is repeated until the maximum value of $\sigma$ is reached.

\begin{figure}[!b]
\includegraphics[scale=0.4]{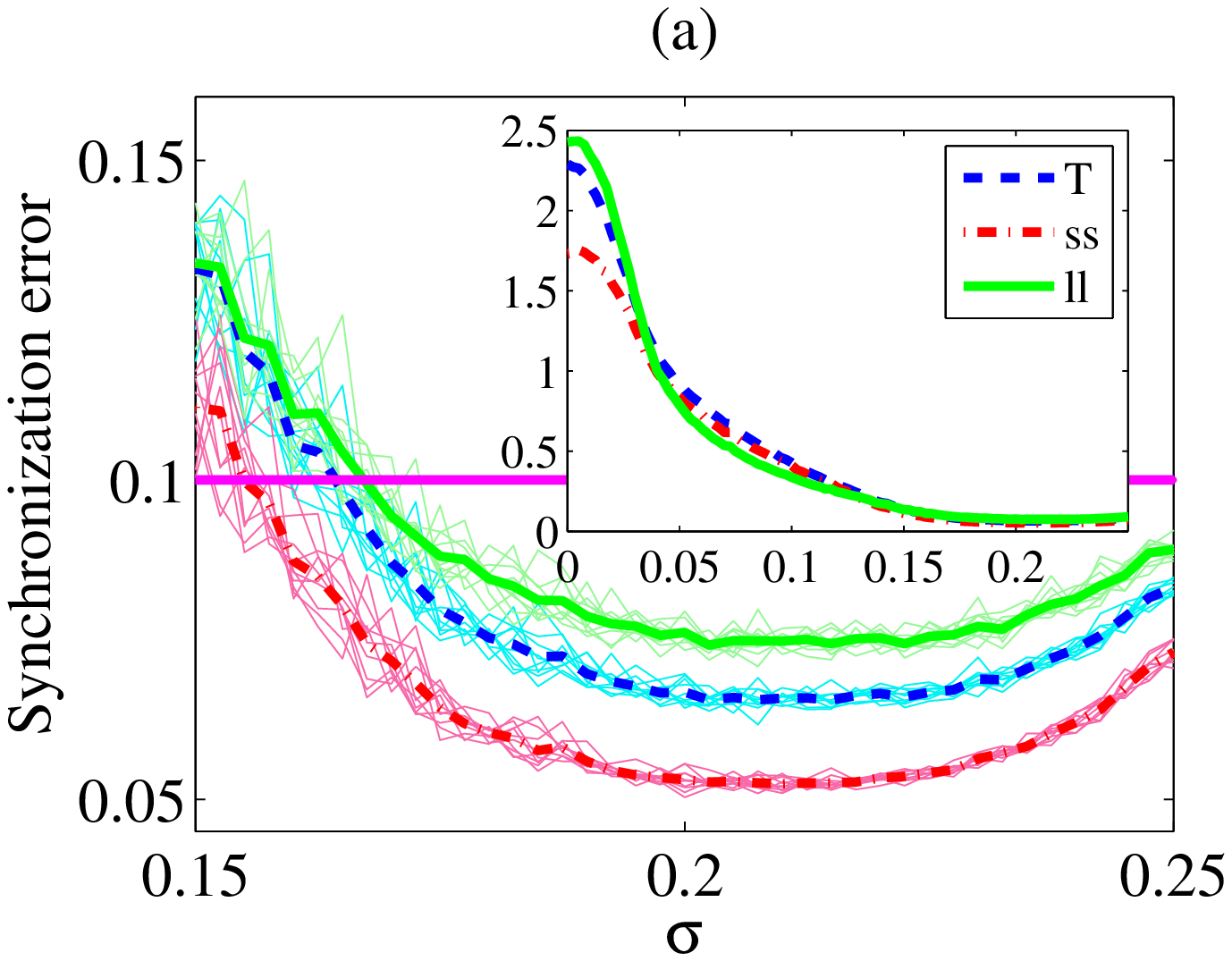}
\includegraphics[scale=0.4]{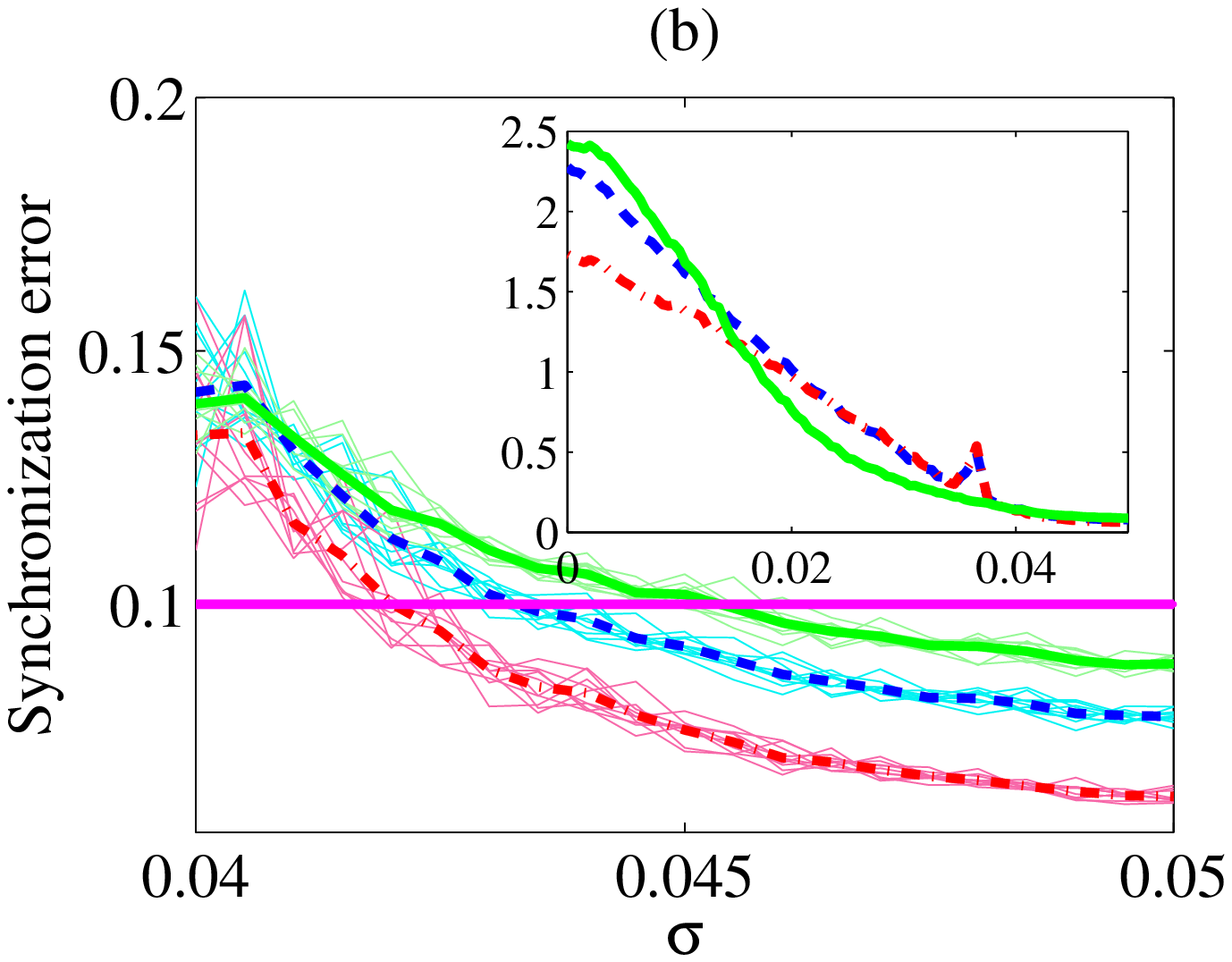}
\caption{\label{fig7} (Color online) {\bf Experimental results: synchronization error as a function of the coupling strength.} (a)  Synchronization error (see main text for a definition) for a network of oscillators coupled through the $x$ variable (class III system) for whole time series $T$ (blue, dark gray, middle curves), time windows $ss$ where all oscillators exhibit $S$ dynamics (red, gray, lower curves), and time windows $ll$ where all oscillators have $L$ dynamics (green, light gray, upper curves). Insets: the curves zoomed around the smallest synchronization error achieved in the system, the results obtained from all $10$ individual realizations (dotted lines) superimposed on the averages (continuous lines). (b) Analogous curves for a network of oscillators coupled through the $y$ variable (class II system).}
\end{figure}

Figure \ref{fig7} shows the synchronization error of the whole network calculated as $\frac{2}{N\cdot (N-1)}\sum_{i<j} |x_i - x_j|$, where the normalizing factor corresponds to the total number of oscillator pairs in the network. We repeat the experiment from $10$ different initial conditions and compute the average synchronization error across the realizations. Whatever the initial condition, the $6$-oscillator network exhibits strongly intermittent dynamics which manifests itself by intermittent switches between $S$ and $L$ in every oscillator node. Therefore, we compute the synchronization error in three different ways: I) the total error of the whole time series regardless of visited states, $T$, blue (dark gray) dashed line II) the error during the time windows where all oscillators exhibit $S$ dynamics, red (gray) dash-dotted line, and III) the error during time windows, where all oscillators represent $L$ dynamics, green (light gray) continuous line. The synchronization errors for these three cases when the coupling is introduced through the $x$ variable is shown in Fig. \ref{fig7} (a). In the topology under study, the ratio of the largest to the smallest nonzero eigenvalue of the Laplacian matrix is $\lambda_N/\lambda_2 =  2.519$, that is smaller than the ratio of the largest to the smallest zeros of the MSF of $L$  $\nu_2/\nu_1=3.97$ (and obviously smaller than the conspicuously larger analogous ratio for $S$). There is thus a range of values of the coupling strength $\sigma$ for which all  eigenmodes are predicted to be in the stability region (i.e., $\sigma \lambda_2>\nu_1$ and $\sigma \lambda_N<\nu_2$ hold simultaneously), and the system is synchronizable. Indeed, we observe that, when $\sigma$ increases, complete synchronization of the whole network is achieved around $\sigma=0.165$  \cite{experror}. The experimental system becomes unstable (the oscillations suddenly disappear and the system reaches a fixed point) above $\sigma \simeq 0.25$, so our results are most relevant in relation to the eigenmode corresponding to $\lambda_2$ and the crossing of the first MSF zero (given the $\lambda_N/\lambda_2$ ratio, at the time this eigenmode crosses only slightly the stability region, the other eigenmodes are guaranteed to be stable).
Figure \ref{fig7} (b) shows the same results for the case of $y$-coupling. In this case the system is indeed class II, so again we focus on the crossing of the eigenmode corresponding to $\lambda_2$ beyond the (only) zero of the MSF. The network reaches synchronization for $\sigma>0.0435$ and does not leave the synchronized manifold for larger values of the coupling strength, as predicted by the MSF.

At this point, it is interesting to analyze the influence of the parameter mismatch on the emergence of synchronization. First, we study the effect of parameter uncertainties in the MSFs. All parameters in our system ($\alpha_1$, $\alpha_2$, $\alpha_3$, $\beta$, $\Gamma$, $\gamma$, $\delta$ and $\mu$ in Eq. (\ref{Rossler1})) are resistances, capacitances, and products thereof (see \cite{Pisarchik08b} for the details). The insets of Fig. \ref{fig5} show the effect of the uncertainties in these parameters (taking the resistor tolerance to be $1\%$ and the capacitor tolerance to be $10\%$) on 50 realizations for each possible case ($x$-coupling for $S$ and $L$ in panel (a), and $y$-coupling for $S$ and $L$ in panel (b)). We zoom around the first zero of each MSF, as this is the most relevant region for the comparison with experiment. Deviations from the noiseless case of the order of $20\%$ or higher are frequently seen. Nevertheless, the MSFs for $S$ and $L$ do not overlap despite the parameter fluctuations and therefore are easily distinguishable. This is in a good qualitative agreement with our experiments. The insets of Figs. \ref{fig7} (a)-(b) show the results zoomed around the transition to the synchronized regime, indicating the values of the $10$ individual realizations and their corresponding average, showing a relatively small variability that qualitatively agrees with our numerical predictions on the effects of a parameter mismatch as shown in Fig. \ref{fig5}.

\begin{figure}[!h]
\includegraphics[scale=0.4]{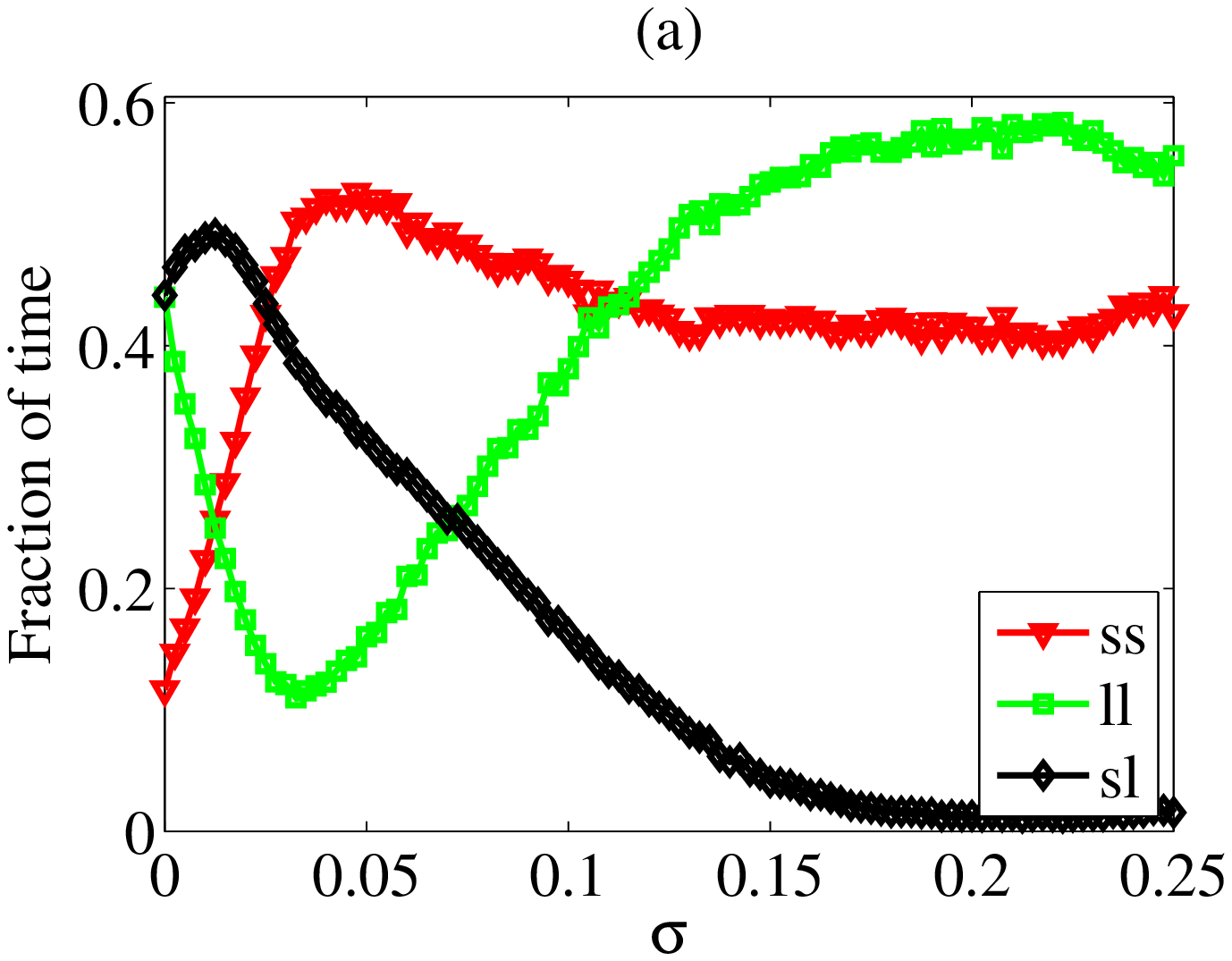}
\includegraphics[scale=0.4]{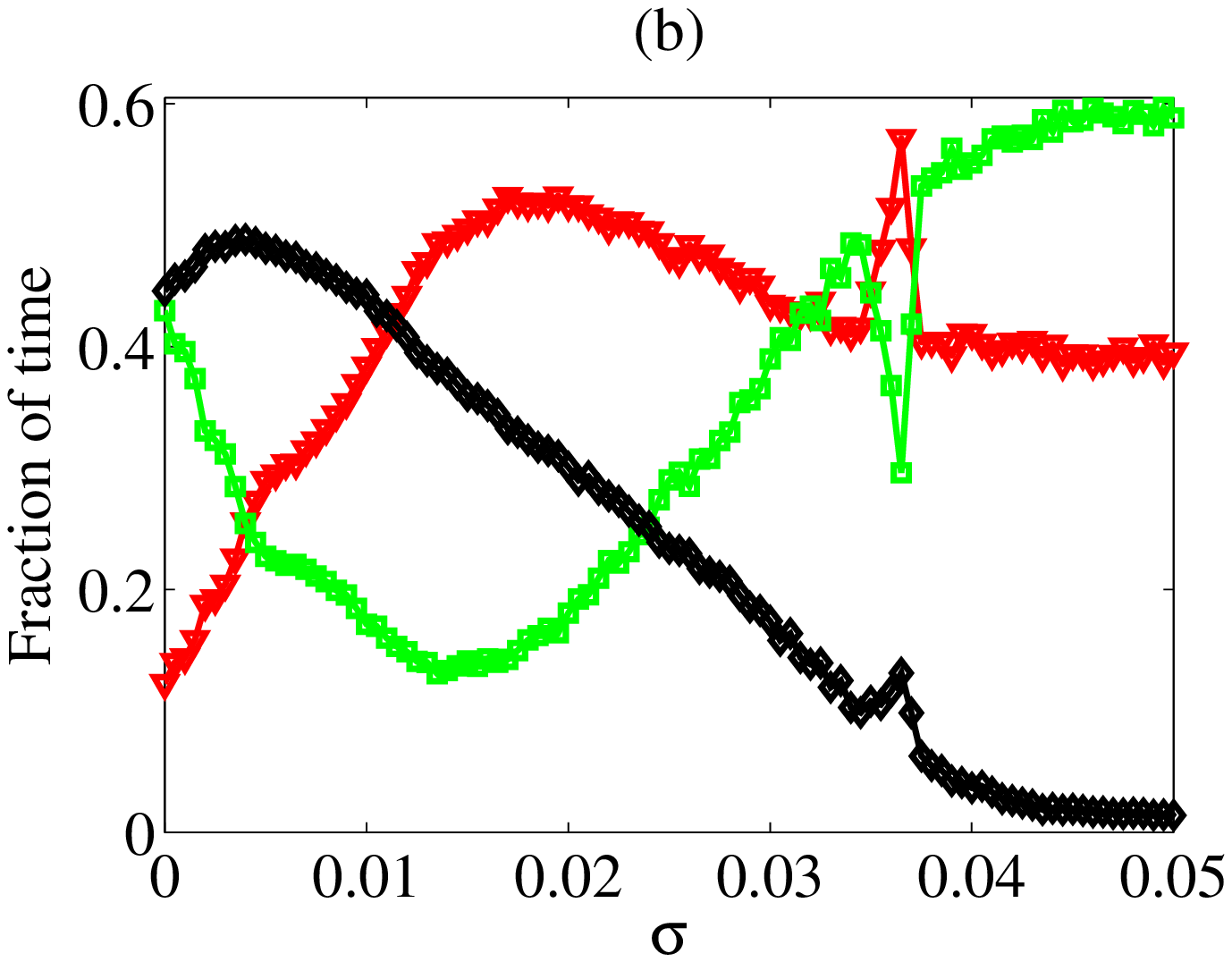}
\caption{\label{fig8} (Color online) {\bf Experimental results: fraction of time that  intermittent system spends in $ss$, red (gray) triangles, $ll$, green (light gray) squares, and $sl$, black (dark gray) diamonds, time windows as a function of $\sigma$.} (a) Network coupled through the $x$ variable and (b) network coupled through the $y$ variable. The results confirm qualitatively the numerically predicted behavior shown in Fig. \ref{fig4}.}
\end{figure}

Finally, Fig. \ref{fig8} shows the fraction of time that the intermittent $I$ system spends in $ss$, red (gray) triangles, $ll$, green (light gray) squares, and $sl$, black (dark gray) diamonds, time windows. One can see the qualitative agreement with the case of two coupled oscillators shown in Fig. \ref{fig4} (a).

\section{Conclusions}	

We have shown how the MSF approach can be used for the analysis of synchronization in ensembles of multistable chaotic systems. To introduce our methodology, we have used a R\"ossler-like oscillator with two coexisting chaotic attractors. The existence of interwoven basins of attraction hinders the  prediction of the system asymptotic behavior in the presence of noise or parameter uncertainties. Under this framework, we have analyzed the synchronization regimes of two bidirectionally coupled chaotic oscillators and showed the enormous complexity of the basins of attraction leading to diverse dynamics of the whole system, as well as the fact that the coexisting attractors became synchronized at different coupling strengths, manifesting various synchronization types for a given coupling strength. After that, we have proposed the use of the MSF arguments for multistable systems, which provided information on synchronizability of a given network of multistable oscillators. Specifically, the MSF shows under what conditions a network of multistable systems can synchronize for a given range of topology spectra and coupling strengths, whatever might be the attractor dynamics to which different oscillators become locked in the presence of intermittency. Even though in this work we have focused on chaotic dynamics, it is important to point out that our approach, as based on a MSF reasoning, can also be used to study synchronization of multistable periodic systems (a MSF approach to monostable periodic, as well as chaotic, dynamics appears for example in  \cite{Barahona2002}). Finally, we have experimentally demonstrated the feasibility of the MSF approach with a network of oscillating circuits in a heavily intermittent regime, and shown that the predictions, as in the case of monostable systems, affected by small uncertainties, were nonetheless very useful from a qualitatively point of view, showing the robustness of the proposed methodology to noise and parameter mismatch.

The proposed methodology is of a special interest for a stability analysis of synchronization in multistable systems during intermittency, since any monostable approach to synchronization is bound to fail in that regime, and knowing the coupling strength or the topological modifications required to maintain complete synchronization under any possible attractor dynamics is especially useful in that scenario. The MSF arguments in the context of multistable systems provide a generic tool to understand complete synchronization of {\it real} multistable systems such as those occurring in laser physics \cite{arecchi1982}, genetics \cite{Vetsigiana09} or cell signaling \cite{Angeli04}.

\section{Acknowledgements}

Authors acknowledge J. Aguirre, D. Papo and P.L. del Barrio for fruitful conversations and the support of MINECO
(FIS2012-38949-C03-01 and FIS2013-41057). R.S.E. acknowledges Universidad de Guadalajara, CULagos (Mexico) for financial support (PIFI 522943 (2012) and Becas Movilidad 290674-CVU-386032).

\section{Appendix - Computation of Lyapunov exponents}

The computation of Lyapunov exponents of coupled R\"ossler-like systems throughout this paper has been performed using the well established method first proposed and justified theoretically in Refs. \cite{80BGGS}. The reader may, however, raise the objection that the equations of motion of the system under study are given by a non-smooth vector field, whose first derivate presents a discontinuity, and therefore the validity of our results is not clear at this point. Indeed, the equation of motion for the $z$ variable in Eq. (\ref{Rossler1}) contains a piecewise linear function $g(x)$, whose derivative $g'(x)$ is such that $\lim_{x\to3^-} g'(x) \neq \lim_{x\to3^+} g'(x)$. Even if the mathematical problem of the existence of Lyapunov exponents for a sufficiently general dissipative dynamical system remains unsolved to this very day \cite{08OY}, from a more practical perspective, the difficulty of computing quantities that are based on the linearized variational equations of a dynamical system whose Jacobian matrix is undefined in the phase-space plane $x=3$ is already apparent. So there are grounds for this objection, even if in practice the Lyapunov spectrum of our systems converge to well defined asymptotic values. On the other hand, a more practically minded reader would point out that the plane $x=3$ is a zero Lebesgue measure set, and that even in the discretized world of numerical analysis the chances that we get states codified with double precision such that $x=3.000000\ldots$ are very small. The argument is that if indeed such states ever occur, their contribution to the Lyapunov exponents (which are effectively computed as long time averages across phase-space orbits) will be negligible. We will see that this latter position turns out to be vindicated by a close examination of the issue.

\begin{figure}[!b]
\includegraphics[scale=0.40]{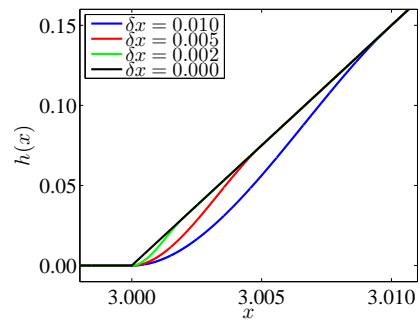}
\caption{\label{figA1} (Color online) {\bf Function $g(x)$ around $x=3$ (black line, $\delta x = 0$), and function $h(x)$ for three different $\delta x$ values (blue line, $\delta x = 0.010$; red line, $\delta x = 0.005$; green line, $\delta x = 0.002$).} For sufficiently small $\delta x$, $h(x)$ can be made as close as needed to $g(x)$ without discontinuities in the first derivative. Higher degree polynomials could be used in a completely analogous manner to approximate $g(x)$ with a higher number of continuous derivatives at $x=3$ and $x = 3 +\delta x$.}
\end{figure}

To avoid the above mentioned non-smoothness, the idea of using a slightly modified dynamical system, where the discontinuity at $x=3$ is bypassed at sufficiently small scales in such a way that it does not affect too much the estimation of the Lyapunov spectrum, is one that naturally comes to mind. As a matter of fact, our choice of a dynamical system was ultimately determined by the fact that the piecewise R\"ossler system is known for its robustness and experimental accessibility when implemented in electronic circuits. There, the discontinuity in the slope of $g(x)$ is obtained by using a semiconductor diode placed in a voltage divider, whose $I-V$ characteristics is supposed to be such that $I>0$ for $V=V_d$ and $I=0$ for $V<V_d$. But of course, this is known to be a crude simplification of the diode behavior, as statistical physics models considering the transport of charge carriers across the depletion layer at the semiconductor $p-n$ junction result in a smooth function in the macroscopic $I-V$ behavior. One possibility would be to include equations based on the Shockley diode law or more realistic mathematical models of a diode, and obtain the relevant $g(x)$ of what would be a considerably  more complicated smooth dynamical system. Instead, we want to explore the possibility of simply replacing $g(x)$ with a perturbed smooth function $h(x)$ defined as follows
\begin{equation}
h(x)=\left\{
\begin{array}{cc}
0 & x\leq 3, \\
p(x) & 3<x\leq 3+\delta x,\\
\mu \left( x-3\right)  & x>3+\delta x,%
\end{array}%
\right.  \label{g}
\end{equation}
where $\delta x \ll 3$. Here, $p(x)$ can be a polynomial of a degree so high to have as many continuous derivatives as one wishes at $x=3$ and $x=3+\delta x$, or a more sophisticated interpolating function. For the sake of simplicity, let us take $p(x)$ to be a third degree polynomial. Four undetermined constants guarantee that we can satisfy four minimal conditions to provide continuity of $h(x)$ and $h'(x)$ at both $x=3$ and $x=3+\delta x$. The resulting polynomial $p(x) = -\frac{\mu}{\delta x^2}(x-3)^3 + \frac{2\mu}{\delta x} (x-3)^2$ is shown in Fig. \ref{figA1} for $\delta x = 0.010$ (blue line), $0.005$ (red line) and $0.002$ (green line). We also include the limiting case $\delta x = 0$ (black line), for which $h(x) = g(x)$.

\begin{figure}[!b]
\includegraphics[scale=0.40]{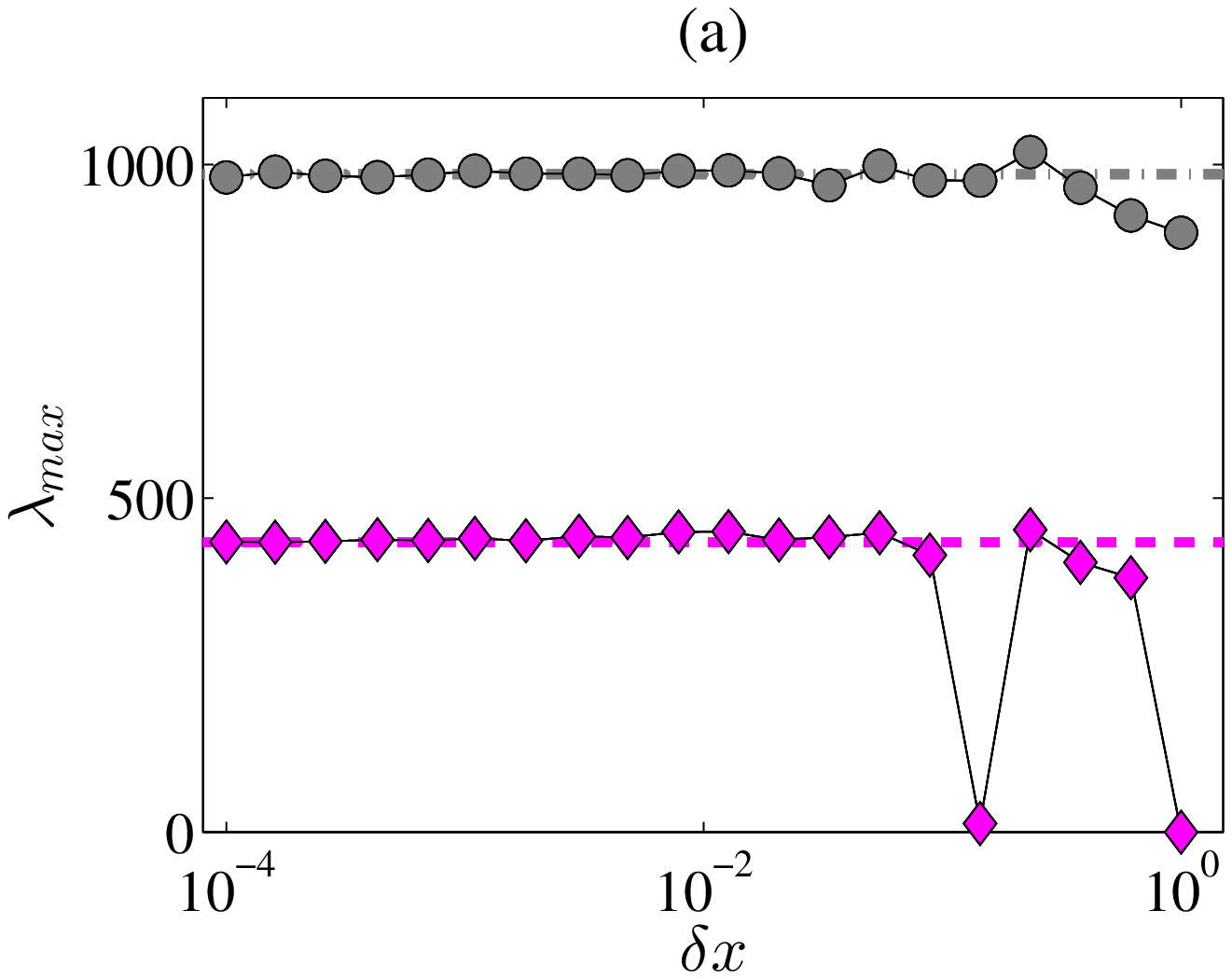}\\
\includegraphics[scale=0.40]{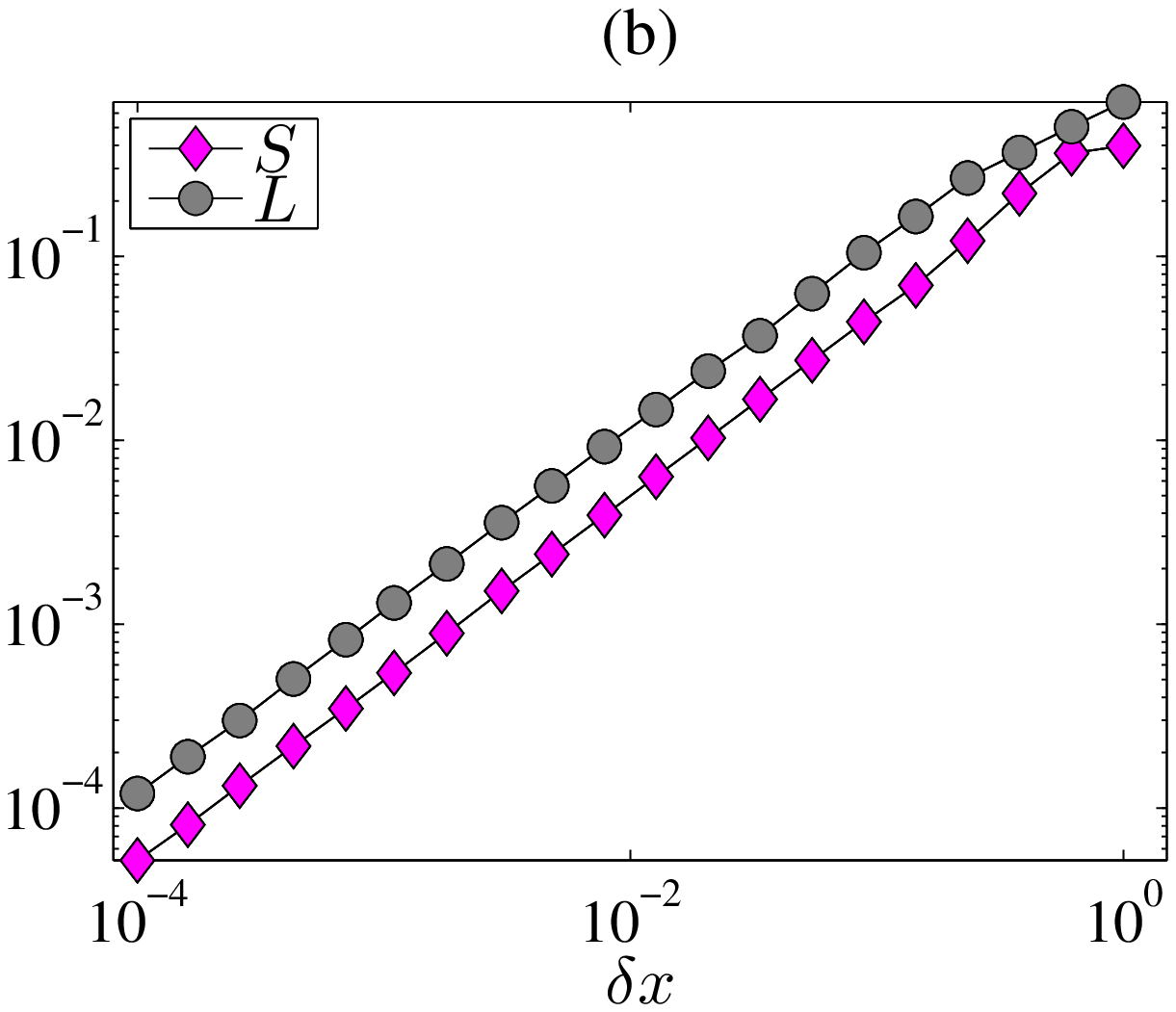}
\caption{\label{figA2} (Color online) {\bf Maximum Lyapunov exponents as a function of $\delta x$, and fraction of phase space points for which $x \in [3,3+\delta x]$.} In (a), the largest Lyapunov exponents corresponding to attractors $S$ (magenta diamonds) and $L$ (grey circles) as functions of $\delta x$. The dashed lines in the background correspond to the estimates obtained with $\delta x = 0$ ($h(x) = g(x)$). In (b), the fraction of phase space points within the interval $[3,3+\delta x]$, where $h(x) \neq g(x)$.}
\end{figure}

Next, we check how the estimated Lyapunov exponents of the R\"ossler oscillator are affected by replacing $g(x)$ with $h(x)$ in the equations of motion. We expect the effect to be smaller and smaller as $\delta x$  decreases. This expectation is indeed matched by the results, as can be seen in the upper panel of Fig. \ref{figA2}. Here, the largest Lyapunov exponents corresponding to attractors $S$ (magenta diamonds) and $L$ (gray circles) are shown. They result from the integration of the system across a time window comprising roughly 30000 cycles. For large $\delta x$, the change in the dynamics significantly affects the Lyapunov exponents, as it is seen quite clearly in the fact that $S$ for some choices of $\delta x$ becomes a regular (i.e. non-chaotic) attractor with a zero maximum Lyapunov exponent. On the other hand, for any choice of $\delta x$ such that $\delta x < 0.01$ all maximum Lyapunov exponent estimates converge, within our accuracy, to the same values, and that is also the case for the rest of the Lyapunov spectrum (not shown here).

The previous results can be simply related to the fact that for small $\delta x$ a small fraction of the phase space points fall inside the region where $h(x)$ differs from $g(x)$. Indeed, we see in  Fig. \ref{figA2} (b) that this fraction decreases as $\delta x$ is reduced as a power law with a characteristic exponent close to $1$. Therefore, replacing $g(x)$ with $h(x)$ with a sufficiently small $\delta x$ solves the issue of the discontinuity in the Jacobian of the system giving Lyapunov exponent estimates independent of the precise values of $\delta x$.

Finally, we consider the possibility of taking the limiting case $\delta x = 0$ ($h(x) = g(x)$), which is tantamount to simply ignoring the discontinuity in the Jacobian. In practice, we use the criterion that if a phase space point happens to be such that $x=3$, we consider the derivative of the $z$ component of the vector field with respect to $x$ to be its left derivative (i.e. the derivative of $g(x)$ is set to $\lim_{x\to3^-} g'(x) = 0$), which is also our choice throughout the paper. The corresponding Lyapunov exponent estimates are shown in Fig. \ref{figA2} (a) as the dashed lines (magenta for $S$, gray for $L$). They coincide with the values obtained by replacing $g(x)$ with $h(x)$ for a sufficiently small $\delta x$. We could also choose to assign as derivative of $g(x)$ at $x=3$ half of the time the left derivative, half of the time the right derivative, or any other reasonable criterion. Ultimately, this choice is inconsequential, as the fraction of points that with $x=3$ is zero for both $L$ and $S$, none of the considered 30 million phase-space points has an $x$ coordinate that is {\it exactly} 3 (in the double precision floating point format). As expected from the results shown in Fig. \ref{figA2} (b), this will also be the case for the slightly perturbed system with $h(x)$ in lieu of $g(x)$ if $\delta x$ is so small that the fraction of points with $x \in [3,3+\delta x]$ times the total number of orbit points considered is considerably smaller than a number on the order of unity. Even if a few points in hundreds of thousands or millions of phase-space points were affected by the discontinuity in the Jacobian or by the small modification in the dynamics introduced by the polynomial $p(x)$, the effect on long time averages along phase-space orbits, such as those Lyapunov exponent estimates are based upon, would be negligible.

In conclusion, in order to avoid the discontinuity in the Jacobian of our dynamical system, we can compute the Lyapunov exponents using a slightly modified dynamical system. Doing this, we must ensure that the modification should be sufficiently small so that the results become independent of the size of the phase-space region whose dynamics is modified and of the modification form (independent of the size of the modified region $\delta x$ and of the degree of the polynomial $p(x)$ in the example above). But then, the resulting Lyapunov exponent estimates coincide with those obtained from the original dynamical system.


\begin{thebibliography}{}
\bibitem{detectcriterion}
We classify the dynamics as corresponding to $S$ or $L$ based on counting the number of local maxima in the $x$ variable along every semi-cycle for which $x>0$, which is systematically larger than one for $L$ and exactly one for $S$. In the presence of intermittency, however, the system may jump from $S$ to $L$, or the other way around, while $x>0$, which compromises the quality of the identification of $ss$, $sl$ and $ll$ time windows in Fig. \ref{fig4}.
\bibitem{experror} The minimum value of the synchronization error, which never reaches zero due to the intrinsic noise of the electronic circuits, is  $0.1$ Vpp, and it is determined by the error of two unidirectionally coupled R\"ossler-like circuits in the limit of strong coupling, as explained in \cite{Aguirre14}.
\end{thebibliography}
\end{document}